\documentclass[longauth]{aa} 

\usepackage{graphicx}
\usepackage{color}
\usepackage{txfonts}
\newcommand{\mum}{{$\mu$m}}
\newcommand{\mjy}{{$\mu$Jy}}

\newcommand{\ha}{H$\alpha$}

\newcommand{\hb}{H$\beta$}

\newcommand{\oiii}{[O\,{\sc iii}]}

\begin{document}

   \title{ MIRI Deep Imaging Survey (MIDIS) of the Hubble Ultra Deep Field\thanks{Based on results from the MIRI European Consortium Guaranteed Time Observations, program 1283} }

   \subtitle{Project description and early results for the galaxy population detected at 5.6~\mum }

   \author{G\"oran \"Ostlin
          \inst{1}
          \and
          Pablo G. P\'erez-Gonz\'alez\inst{2}
          \and
          Jens Melinder \inst{1}
          \and\ 
          Steven Gillman \inst{3,6}
          \and
          Edoardo Iani \inst{4}
          \and
          Luca Costantin \inst{2}
          \and
          Leindert A. Boogaard \inst{5}
          \and
          Pierluigi Rinaldi \inst{27,4}
          \and          Luis Colina \inst{2}
          \and Hans Ulrik N\o rgaard-Nielsen \inst{3,29}
          \and
          Daniel Dicken  \inst{7}
          \and
          Thomas R. Greve \inst{3,6,8}
          \and Gillian Wright \inst{7}
          \and Almudena Alonso-Herrero \inst{25}
          \and Javier Alvarez-Marquez \inst{2}
          \and Marianna Annunziatella \inst{2}
         \and  Arjan Bik \inst{1}
         \and  Sarah E.I. Bosman \inst{5,9}
         \and  Karina I. Caputi \inst{4,6}
         \and  Alejandro Crespo Gomez \inst{2}
         \and  Andreas Eckart \inst{10}
         \and  Macarena Garcia-Marin \inst{11}
         \and  Jens Hjorth \inst{12}
         \and  Olivier Ilbert \inst{13}
         \and  Iris Jermann \inst{3}
         \and  Sarah Kendrew \inst{11}
         \and  Alvaro Labiano \inst{25,26}
          \and Danial Langeroodi \inst{12}
         \and  Olivier Le Fevre \inst{13,29}
        \and  Mattia Libralato \inst{14,28}
         \and  Romain A. Meyer \inst{15,5}
         \and  Thibaud Moutard \inst{16}
         \and  Florian Peissker \inst{10}
         \and  John P.  Pye \inst{17}
         \and  Tuomo V. Tikkanen \inst{17}
         \and  Martin Topinka \inst{18}
         \and  Fabian Walter \inst{5}
        \and   Martin Ward \inst{19}
        \and   Paul van der Werf  \inst{20}
        \and Ewine F. van Dishoeck  \inst{20}
        \and Manuel G\"udel \inst{21,22}
        \and Thomas Henning \inst{5}
        \and Pierre-Olivier Lagage \inst{23}
        \and Tom P. Ray \inst{18}
        \and Bart Vandenbussche \inst{24}
          }

   \institute{Department of Astronomy, Oskar Klein Centre, Stockholm University,
              AlbaNova University Center, 10691 Stockholm, Sweden\\
              \email{ostlin@astro.su.se}
         \and
              Centro de Astrobiología (CAB), CSIC-INTA, Ctra. de Ajalvir km 4, Torrejón de Ardoz 28850, Madrid, Spain
        \and 
        DTU Space, Technical University of Denmark, Elektrovej 327, 2800 Kgs. Lyngby, Denmark
        \and
        Kapteyn Astronomical Institute, University of Groningen, PO Box 800, 9700 AV Groningen, The Netherlands
        \and
        Max-Planck-Institut für Astronomie, Königstuhl 17, 69117 Heidel- berg, Germany
        \and
        Cosmic Dawn Centre (DAWN), Copenhagen, Denmark
        \and 
        UK Astronomy Technology Centre, Royal Observatory Edinburgh, Blackford Hill, Edinburgh EH9 3HJ, UK
        \and
        Department of Physics and Astronomy, University College London, Gower Place, London WC1E 6BT, UK
        \and Institute for Theoretical Physics, Heidelberg University, Heidelberg, Germany
 \and        I. Physikalisches Institut der Universität zu Köln, Zülpicher Str. 77, 50937 Köln, Germany
 \and European Space Agency, Space Telescope Science Institute, Baltimore, MD, USA
\and DARK, Niels Bohr Institute, University of Copenhagen, Jagtvej 155A, 2200 Copenhagen, Denmark
\and Aix Marseille Université, CNRS, LAM (Laboratoire d’Astrophysique de Marseille) UMR 7326, 13388 Marseille, France
\and AURA for the European Space Agency (ESA), Space Telescope Science Institute, 3700 San Martin Drive, Baltimore, MD, USA
\and Department of Astronomy, University of Geneva, Chemin Pegasi 51, 1290 Versoix, Switzerland
\and European Space Agency (ESA), European Space Astronomy Centre (ESAC), Camino Bajo del Castillo s/n, 28692, Villaneuva de la Canada, Madrid, Spain
\and School of Physics \& Astronomy,  Space Park Leicester, University of Leicester, 92 Corporation Road, Leicester LE4 5SP, UK
\and        Dublin Institute for Advanced Studies, Astronomy \& Astrophysics Section, 31 Fitzwilliam Place, Dublin 2, Ireland
\and Centre for Extragalactic Astronomy, Durham University, South Road, Durham DH1 3LE, UK
\and Leiden Observatory, Leiden University, PO Box 9513, 2300 RA Lei- den, The Netherlands
\and University of Vienna, Department of Astrophysics, Türkenschanzstrasse 17, 1180 Vienna, Austria
\and
Institute of Particle Physics and Astrophysics, ETH Zürich, Wolfgang-Pauli-Str 27, 8093 Zurich, Switzerland
\and AIM, CEA, CNRS, Université Paris-Saclay, Université Paris Diderot, Sorbonne Paris Cité, 91191 Gif-sur-Yvette, France
\and Institute of Astronomy, KU Leuven, Celestijnenlaan 200D bus 2401, 3001 Leuven, Belgium
\and 
Centro de Astrobiolog\'{\i}a (CAB), CSIC-INTA, Camino Bajo del 
Castillo s/n, E-28692 Villanueva de la Ca\~nada, Madrid, Spain
\and
Telespazio UK for the European Space Agency (ESA), ESAC, Camino Bajo del Castillo s/n, 28692 Villanueva de la Ca{\~n}ada, Spain 
\and 
Steward Observatory, University of Arizona, 933 North Cherry Avenue, Tucson, AZ 85721, USA
\and INAF - Osservatorio Astronomico di Padova, Vicolo dell’Osservatorio 5, Padova I-35122, Italy
\and Deceased
  }

   \date{Submitted Jul 30, 2024}


  \abstract
   {The recently launched James Webb Space Telescope (JWST) is opening new observing windows on the distant universe. Among JWST's instruments, the Mid Infrared Instrument (MIRI) offers the unique capability of imaging observations at wavelengths $\lambda > 5\mu$m. This enables unique access to the rest frame near infra-red (NIR, $\lambda \ge 1$\mum) emission from galaxies at redshifts $z>4$ and the visual ($\lambda \gtrsim 5000$\AA) rest frame for $z>9$. We here report on the guaranteed time observations (GTO) from the MIRI European Consortium, of the Hubble Ultra Deep Field (HUDF), forming the MIRI Deep Imaging Survey (MIDIS), consisting of an on source integration time of $\sim41$ hours in the MIRI/F560W (5.6 $\mu$m) filter. 
   To our knowledge, this constitutes the longest single filter exposure obtained with JWST of an extragalactic field as yet. }
   {The HUDF is one of the most observed  extragalactic fields, with extensive multi-wavelength coverage, where (before  JWST) galaxies up to  $z\sim 7$ have been  confirmed, and at $z>10$  suggested, from HST photometry. We aim to characterise the galaxy population in HUDF at 5.6 $\mu$m, enabling studies such as: the rest frame NIR morphologies for galaxies at $z\lesssim4.6$,  probing mature stellar populations and emission lines  in $z>6$ sources, intrinsically red and dusty galaxies, and active galactic nuclei (AGN) and their host galaxies at intermediate redshifts. }
   {We have reduced the MIRI data using the \emph{JWST} pipeline, augmented by in-house custom scripts. We measure the noise characteristics of the resulting image. Galaxy photometry has been obtained, and photometric redshifts have been estimated for sources with available multi wavelength photometry (and compared to spectroscopic redshifts when available). }
   {Over the deepest part of our image  the 5$\sigma$  point source limit is 28.65 mag AB (12.6 nJy), $\sim0.35$ mag better than predicted by the JWST exposure time calculator.
   We find $\sim2500$ sources, the overwhelming majority of which are distant galaxies, but note that spurious sources likely remain at faint magnitudes due to imperfect cosmic ray rejection in the JWST pipeline.
   More than 500 galaxies with available spectroscopic redshifts, up to $z\approx11$ have been identified, the majority of which are at $z<6$. 
   More than 1000 galaxies have reliable photometric redshift estimates, of which $\sim25$ are  at $6<z<12$. 
   The point spread function in the F560W filter has a FWHM of $\approx0.2\arcsec$ (corresponding to $1.4$ kpc at $z=4$), allowing the near infrared rest frame
   morphologies  for the first time to be resolved up to $z\sim4$. As expected, the  light distributions are smoother than at shorter wavelength, and trace the stellar mass distributions.
    Moreover, $>100$  objects with very red  NIRCam vs MIRI (3.6--5.6 \mum\ $>1$ ) colours have been found, indicating dusty or old stellar populations at high redshifts. 
    }
   {We conclude that MIDIS  surpasses pre-flight expectations and that deep MIRI imaging has a great potential for characterizing the galaxy population from cosmic noon to dawn.  }


   \maketitle
%

\begin{figure*}
   \centering
  \includegraphics[scale=0.85,trim=0cm 0 0 0]{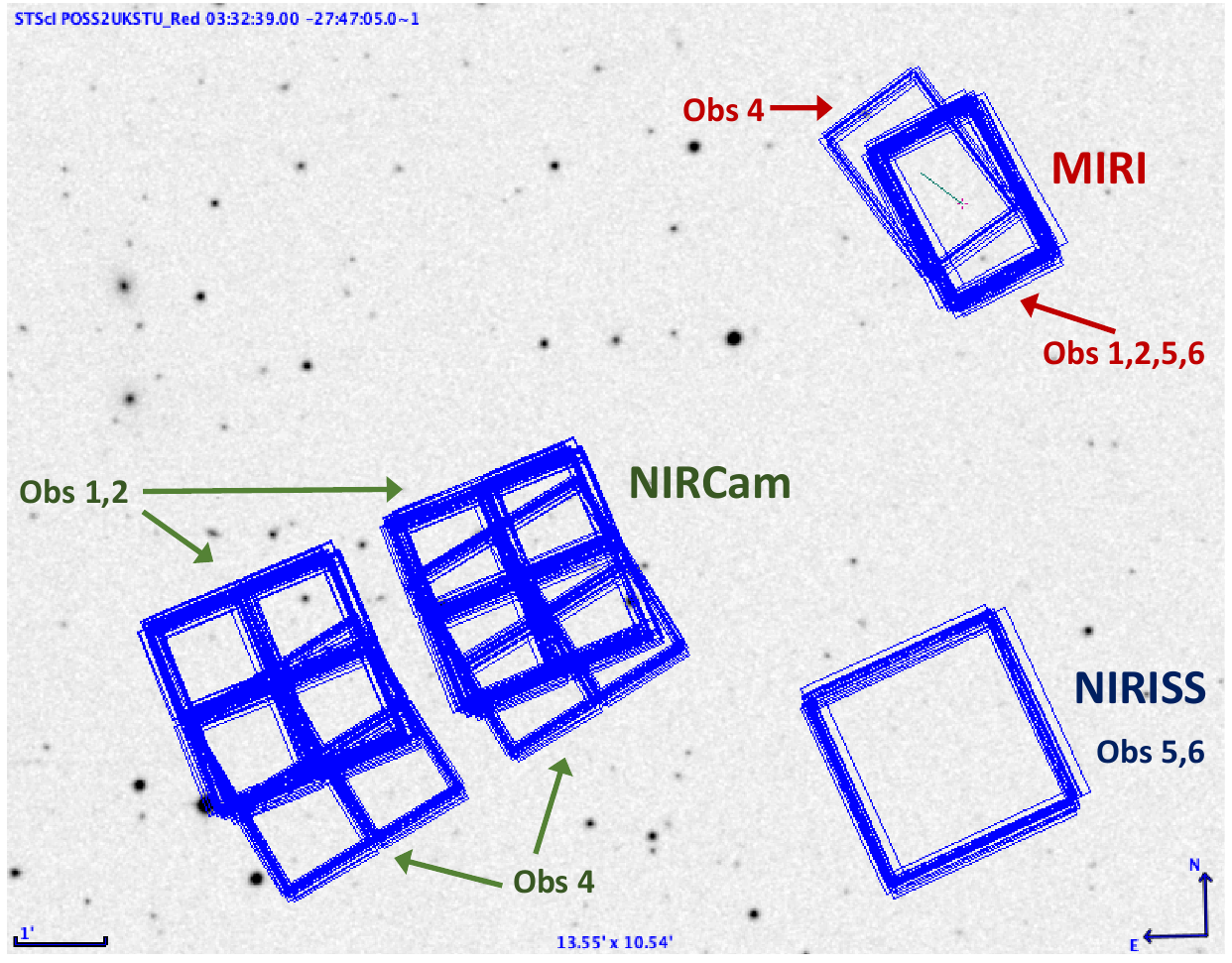}
   \caption{Footprint of all observations obtained in program 1283 overlaid on DSS greyscale image (from APT/Alladin). To the upper right, the MIRI pointings for Observation 1,2,5,6 (thicker rectangle) and 4 (rotated 10$^\circ$ anti-clockwise and offset $\approx34\arcsec$ towards North-East) are shown. Note that the Lyot region of the MIRI imager is not shown.  The parallel fields for NIRISS (P3; Obs 5 and 6) and NIRCam (P2; Obs 1, 2 and 4) 
   are indicated. Obs4 was obtained at a different position angle due to JWST safing event making the schedule slip. }
              \label{aladin}%
    \end{figure*}

\section{Introduction}

The late 1990's marked a revolution in the study of the high-redshift universe. This was made possible through new facilities such as the Hubble Space Telescope (HST), and 8-10m class ground-based telescopes such as the Keck, and its successors, notably the ESO/VLT.  While before, redshifts $z>1$ was mostly the domain of quasi-stellar objects (QSOs) and radio galaxies, breakthroughs such as the Lyman break technique \citep[implying dropouts in the bluer filters due to a redshifted Lyman break]{sh92,steidel96a}, more sensitive narrow-band surveys for Lyman $\alpha$ (Ly$\alpha$) emitters \citep{hu98}, and not the least the advent of the Hubble Deep Field \citep[HDF,][]{williams1996},  dramatically changed the picture \citep{steidel96b}. The HDF was an ambitious undertaking by imaging an apparently blank and almost starless field at high Galactic latitude with HST in several filters with exposure times amounting to 30--40 hours per filter in U, B, V, and I. 
The HDF was found to be anything but blank, containing thousands of distant galaxies, and triggered a wealth of high-z studies. Spectroscopic campaigns confirmed most galaxies in the HDF to be at high redshifts, and dropout and photometric redshift techniques could show with high probability that many sources too faint for ground-based spectroscopy were likely to reside at very high redshifts. The HDF was followed over the years by several other deep field campaigns from the ground, and notably with the HST, for instance HDF-south \citep{williams2000},  GOODS \citep{giavalisco04}, and the Hubble Ultra Deep Field \citep[HUDF, ][]{beckwidth06} which is contained within the GOODS south (GOODS-S) field. Such deep field studies aim at exploring when galaxies first formed and how the have evolved over cosmic time.

At redshifts $z\gtrsim4$ the absorption by neutral hydrogen in the intergalactic medium (IGM) efficiently suppresses any emission bluewards of Ly$\alpha$, meaning that 'dropouts' would mainly trace the Ly$\alpha$ break (1216\AA) rather than the Lyman discontinuity (912 \AA). With optical-only observations this originally limited dropout studies with HST to $z<6$.
From 1997, near infra-red (NIR) capability was added to HST by the NICMOS instrument \citep{thompson01}, which though had a very small field of 
view, but with the advent of HST Wide Field Camera 3 (installed during HST servicing mission 4 in 2009), with its near IR channel operating up to 1.6 $\mu$m
the field again exploded \citep[e.g.][]{oesch10}. Deep observing programs such as CANDELS \citep{koekemoer11} and the HST Frontier Fields \citep{lotz17} led to the detection of many new galaxy candidates at even higher redshifts. The deepest field observed with HST to date remains the HUDF, of which the fraction with the deepest WFC3/IR observations forms the eXtremely Deep Field \citep[XDF, ][]{illingworth13}.

At longer wavelengths, significant capability beyond 2.2 $\mu$m (the K-band, the   reddest band practically useful for deep ground-based surveys) was added by the Spitzer Space Telescope from 2004 and onwards, though it lacked the spatial resolution of HST \citep{fazio04a} and therefore was susceptible to source confusion at faint fluxes. Notably the GOODS areas have been extensively imaged with the IRAC instrument (having broad band filters centered at 3.6, 4.5, 5.8 and 8 \mum) onboard Spitzer. 
The first two filters of IRAC (Ch1 \& Ch2) were observed also during the extended warm phase of the Spitzer mission resulting in exposure times of $\sim200$h in the HUDF, but the two longer wavelength filters (Ch3 \& Ch4) were not observable in the warm mission phase. Hence IRAC imaging of the HUDF is significantly shallower at $\lambda > 5$ \mum\ \citep{stefanon21}. The observations presented in this paper (MIRI with filter F560W) have similar band-pass to IRAC/Ch3, allowing the increase in performance with JWST/MIRI to be quantified. 

The quest for the highest redshift galaxies is ultimately about finding the first galaxies that appeared in the universe. Following the Big Bang and the subsequent recombination (when the Cosmic Microwave Background (CMB) radiation was released at $z\sim1100$) the universe was neutral but with little structure. As density inhomogeneities grew under gravity, primeval stars and galaxies formed and the ionising output from early galaxies and possibly active galactic nuclei (AGN) would eventually be large enough to ionize the intergalactic medium (reionization). Studies of QSO absorption lines \citep[i.e. finding the Gunn-Peterson trough,][]{gp} indicate that this process was largely complete around $z\sim 6$
\citep{becker01, fan06}, but likely extended down to $z\sim5.3$ \citep{bosman22}. Hence, redshifts  $z\gtrsim6$ is commonly referred to as the epoch of reionization (EoR).  Measurements of the CMB Thompson scattering optical depth from the Planck satellite indicate that reionization was half complete, corresponding to an ionised fraction of $x_e = 50\%$, at $z\gtrsim8$, and $x_e<10\%$ at $z\gtrsim 10$ \citep{adam16, pagano20}.

Already in the 1990s, at the times when HST was delivering its first deep images, the need for a successor operating at infra-red (IR) wavelengths was identified, as galaxies at $z>7$ would become invisible at optical wavelengths (due to the Lyman break).
Hence the Next Generation Space Telescope (NGST) project began to be developed for an IR-optimized observatory, and in 2002, NASA renamed it the James Webb Space Telescope (JWST), which was eventually launched on Dec 25, 2021. 

The JWST features four scientific instruments, MIRI, NIRCam, NIRISS and NIRSpec. While the latter three are optimised for imaging and spectroscopy in the near IR, MIRI, the Mid Infrared Instrument \citep{rieke15,wright15} is (as the name implies) optimised for the mid IR (MIR) and uniquely probes the 5--28 \mum\ range in imaging, spectroscopy and coronography.

The launch of JWST, originally (and casually) conceived as 'the first light machine' has opened up a new window for the high-z universe. By extending the red cutoff to much longer wavelengths than HST, higher redshifts can be probed by the imaging dropout technique. NIRCam covers $\lambda\le 5\mu$m with many filters and could in principle detect dropouts to $z\gtrsim30$ (should such galaxies exist and be bright enough).

However, even for a young galaxy, most of the stellar mass is carried by stars whose spectral flux density ($f_\nu$) peaks at rest frame $\lambda_{\rm rest}>0.5\mu$m, which for $z>9$ is the domain of MIRI. Moreover, as evolved stellar populations have redder colours (and reduced flux at $\lambda<0.5\mu$m) the ability to detect an evolved stellar population underlying a younger one increases with wavelength, and MIRI data would therefore be vital for characterizing the stellar populations and star formation histories of galaxies at $z>6$. Moreover, at $z>6.6$, H$\alpha$, the prime probe of star formation activity in galaxies
and which is strong in AGN, shifts into the MIRI range, again making it a vital instrument for studying star formation in galaxies \citep[e.g.][]{rinaldi23}, and supermassive black holes in the EoR \citep[e.g.][]{bosman24}.

Longer wavelengths are also less affected by dust extinction, and massive starbursts are known to have large amounts of extinction (both at low and high-$z$) and the unique ability of MIRI to probe the rest frame NIR ($\lambda>1\mu$m) for $z>4$, and Paschen $\alpha$ for $z>2$ offers unique opportunities to probe such galaxies \citep[e.g. GN20, ][]{colina23}. Many such massive galaxies also contain an AGN whose hot dust torus emission tends to dominate the SED at $\lambda_{\rm rest} >2\mu$m which for $z>1$ can be uniquely probed by MIRI \citep{lyu22,perez24a}.

These considerations led the MIRI European Consortium (MIRI-EC) to allocate $\sim 64$ hours of their guaranteed time observations (GTO), for deep imaging 
of the HUDF/XDF in with MIRI in filter F560W, forming the MIRI Deep Imaging Survey (MIDIS, program 1283). The F560W filter was selected as it (based on the pre-flight sensitivity estimates) reached the faintest flux limit in a given time of the MIRI filters, and with the intent to seed future GO deep imaging projects of the HUDF with longer wavelength MIRI filters.

Since the commissioning of JWST, a number of studies based on Early Release Observations (ERO), Early Release Science (ERS), GTO, and General Observer (GO) programs have demonstrated the power of JWST to probe galaxies in the EoR. Notable are the results from NIRCam, where ERO/ERS/GTO-based studies provided many candidates with probable redshifts $z>9$ \citep[see][]{harikane23, harikane24,perez23}, and the results from the large GTO program JADES \citep{robertson23, rieke23}. MIRI was less used in the ERO/ERS programs, but its potential for characterizing galaxies in the EoR has been demonstrated \citep[e.g.][]{papovich23}. The EoR is though just the tip of the iceberg, and both NIRCam and MIRI can revolutionise the study of vastly more galaxies from the cosmic noon ($z\gtrsim1$) to the EoR, by probing rest-frame optical and near-IR emission where HST is severely compromised by its red cutoff at $\lambda\sim1.6$\mum, and the decomissioned Spitzer/IRAC by its wider point spread function (PSF) and lower sensitivity. This has already been demonstrated in several studies. 

In this paper, we present an overview of the MIRI Deep Imaging Survey (MIDIS), a MIRI European Consortium (MIRI-EC) GTO project (program ID 1283), imaging the HUDF/XDF with MIRI at 5.6\mum, forming the longest exposure of a single extragalactic field with MIRI obtained as yet \citep[see][]{casey23}. In parallel deep NIRCam imaging and NIRISS slitless spectroscopy was obtained for two adjacent fields with deep HST coverage. 
The parallels were chosen to enable selection of: $z>6$ galaxy candidates through the drop-out technique (NIRCam); and intermediate-$z$ emission line and continuum bright galaxies (NIRISS).  The MIDIS project has already resulted in a number of papers 
and more are in preparation, as described below.

The outline of the paper is as follows: in section 2 we describe the observations, in section 3 we describe the reductions and quantify the photometric depth; section 4 mentions the ancillary data used, section 5 describe photometric measurements including a comparison with Spitzer, section 6 describes the photometric redshift estimation, while section 7 give examples of results MIDIS, and section 8 describes how to access the deep MIRI F560W images and photometric catalogues of MIDIS. Section 9 contains a summary and conclusions.

Throughout the paper, all magnitudes quoted are in the AB system.
We assume a cosmology with $H_0=70$\,km\,s$^{-1}$\,Mpc$^{-1}$, $\Omega_m=0.3$, $\Omega_\Lambda =0.7$. 
   

\begin{figure}
   \centering
  \includegraphics[scale=0.7,trim=0cm 0 0 0]{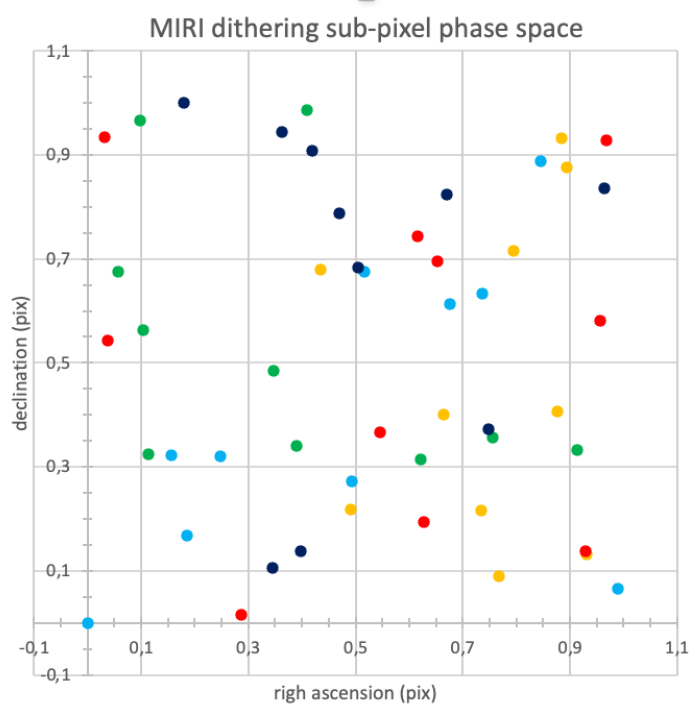}
   \caption{MIRI imaging subpixel phase space sampling near detector centre, where the different colors refer to the different observations (1,2,4,5,6). The x-axis shows $|{\rm frac}(x_i-x_1)| $ (i.e. the fractional pixel shift of the i'th dither position with respect to the 1st dither position in Obs 1) and the corresponding quantity on the y-axis. The pixel scale for MIRI is 0.11\arcsec/pixel.}
              \label{drizzling}%
    \end{figure}

\begin{figure*}
   \centering
  \includegraphics[scale=0.33]{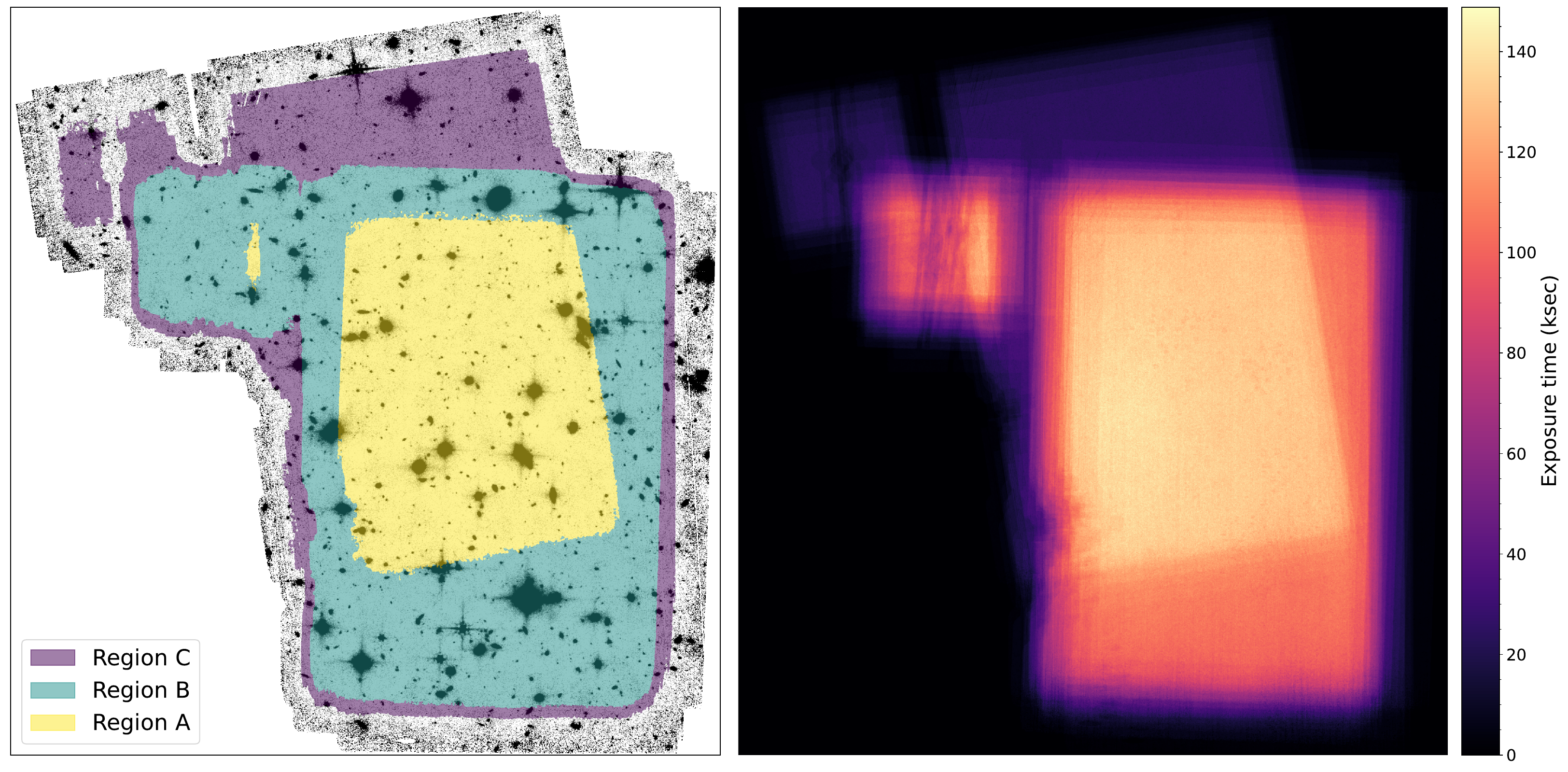}
   \caption{Left: The F560W image with linear intensity scaling. Colored patches show regions where we have estimated the noise (see Table~\ref{depthtable}) with different depths: the deepest area (A), where obs 1-2, 4-6 overlap (yellow). The deep area (B), where obs 4 does not overlap fully with the others (green). Outside these areas, there is coverage at less depth, notably the NE extension which only has observations from obs 4. The area with at least 7h of combined integration is denoted C (purple). The orientation is in the detector plane (x,y). Right: Exposure time map.}
              \label{image-exposuremap}%
\end{figure*}

\begin{figure*}
   \centering
  \includegraphics[scale=1.02,trim=0.42cm 0 0 0]{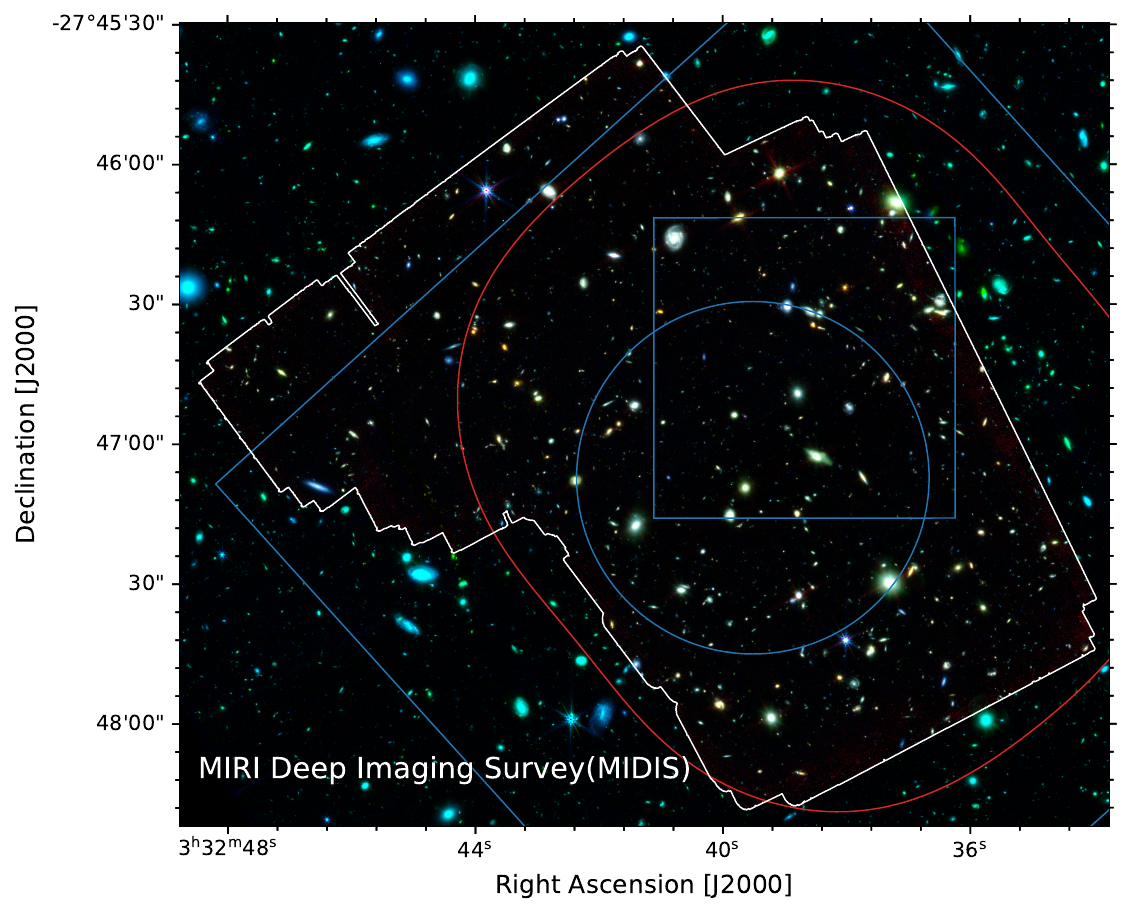}
   \caption{RGB composite of the MIDIS field, with its boundaries outlined by the white contours (the irregular shape of the field is due to the use of two different pointings/PAs and that the nominal FOV has an extra square at the top left where the Lyot coronograph is located). The F560W filter is shown in red, NIRCam F356W in green and F150W in blue \citep[the latter two from JADES,][]{rieke23}. Note that the sources outside the white contour appear blueish/greenish due to the lack of MIRI data there. The thin blue contours outline deep MUSE spectroscopic surveys at 10\,h and 30\, (large and small squares) and the MUSE eXtreme Deep Field (MXDF, circle) reaching up to 140\,h \citep{Bacon2023}, and the red contour outlines the extent of the deep ALMA data from ASPECS at 1.2\,mm and 3\,mm (see \citealt{boogaard24} for details).   }
              \label{rgb}           
    \end{figure*}

\section{Observations}

A single pointing coincident on the XDF was selected for imaging in a single MIRI filter, F560W. Parallel observations were defined for about 63\% of the time with NIRCam (targeting parallel field P2), and 37\% with NIRISS (targeting parallel field P3). Both parallel fields have deep HST imaging \citep{whitaker19}. As a special requirement, we requested  the observations to be executed under low-background conditions. 

The MIRI observations were designed to be split up into six observations (or spacecraft visits), four (Obs 1-4) with NIRCam in parallel and two (Obs 5-6) with NIRISS in parallel. 

For Obs 1-4,  MIRI data acquisition was designed to use the FASTR1 readout mode with 100 groups per integration, 10 integrations per dither position, and 10 dither positions for a total exposure time of 28\,000 seconds (28 ks) per observation. 
For dithering, the CYCLING-MEDIUM pattern was adopted that provides 0.5 pixel (the MIRI imager pixel scale is 0.11\arcsec/pixel) sampling for MIRI imaging. 
This dithering pattern was selected to provide offsets between exposures that were larger than the anticipated sizes of all but the brightest objects in the field.
Observations 1,2,3 and 4 (each consisting of 10 positions) were selected to use starting points 1, 11, 21, and 31, respectively, to have each exposure at a different position. 
In order to further improve the PSF sampling, small (one quarter to one third MIRI pixel) manual offsets were employed at the beginning of each subsequent observation following observation 1 (by using the special requirement OFFSET).  All observations were to be obtained at the same position angle (PA). Obs 1-2 would use NIRCam F115W/F277W in parallel and Obs 3-4 NIRCam F150W/F356W. 

For Obs 5 and 6, NIRISS  was selected as the parallel instrument. For Obs 5 a single 100 group F560W exposure (277.5 s) was obtained (coincident with a parallel NIRISS F115W direct image) at the nominal position, followed by a 9-point dither (again using CYCLING-MEDIUM) with 100 groups and 4 integrations (10065 s, when parallel NIRISS slitless spectroscopy was obtained), and then another single F560W 100 group exposure (again with NIRISS/F115W imaging in parallel). The dither pattern starting point was 41. The same sequence was repeated (using the same dither positions) with NIRISS/F150W in parallel. Finally, the same sequence was repeated with NIRISS/F200W in parallel, but here the initial and final F560W image (where direct NIRISS/F200W images were obtained) had two integrations per exposure (557.8s). The direct images were obtained with no dithers at the nominal position. Hence, for each of the nine points in the cycling pattern we have 12 integrations with 100 groups (a total of 3\,355 seconds per position), and for the nominal position 8 integrations with 100 groups (total 2\,225.6 seconds). 

For Obs 6, the same parameters were used but here using dithering starting point 50, following a 0.0328 arcsec shift in x and y, again to improve the MIRI sampling, and with the NIRISS spectral observations using the grism with orthogonal dispersion direction (GR150C).
Obs 5-6 were to be obtained at the same PA, and with a 5 degree tolerance with respect to Obs 1-4. 

The observations were scheduled for early December 2022, and Obs 5-6, and 1-2 were successfully obtained on Dec 2-6. However, repeated JWST safing events prevented a timely execution of Obs 3-4, and as time slipped and the spacecraft V3 angle was evolving, we  were forced to revise the PA constraints for Obs 3-4, and we added a small spatial offset to ensure better joint coverage of the NIRCam parallel (using filters F150W and F356W) with Obs 1 and 2 (using filters F115W and F277W). In the end, only Obs 4 could be executed, and Obs 3 was postponed to December 2023. In light of this, it was decided to dedicate the MIRI observing time of Obs 3 to F1000W imaging (with NIRCam F150W + F356W in parallel) at the same pointing and PA as for Obs 1, and observations (now relabeled as Obs 7) were executed December 6, 2023. These F1000W observations will be presented in a forthcoming paper (\"Ostlin et. al, 2024, in prep.; see also \citealt{cerberus}).

Hence, for the MIRI/F560W imaging we obtained 5/6 of the requested observations, at 3 different PAs, and at 50 different dither positions, see Table 1. The total net exposure time in F560W amounts to 148\,842 seconds (41.34 hours). To our knowledge, this made it the longest imaging exposure obtained of a single extragalactic field obtained with JWST at the time.

The fact that Obs 4 was offset and rotated with respect to Obs 1-2 means that we cover a larger area, but with less depth in the NE parts. The use of different starting positions for the 5 observations (enabled by the OFFSET special requirement) with
non-integer pixel shifts, and the use of 3 different PAs means that the subpixel phase space is quite uniformly populated, see Fig.~\ref{drizzling}, which is valid near 
the field centre (further away, the relative positions are modulated by the geometric distortion of the MIRI imager).

The total exposure times in the parallel observations are slightly shorter (by 2\% for NIRCam, 27593s per Obs; and 5\% for NIRISS, with 2147s for imaging and 28699s for grism spectroscopy, split over the three settings). 

The footprints of the MIRI primary observations and the parallel NIRCam/NIRISS observations are shown in Fig.  \ref{aladin}. In Fig. \ref{image-exposuremap} we show 
the image  and exposure map for the mosaic of all F560W 
observations (see section 3). Due to the use of different pointings and PAs the mosaic boundary has a complex shape.

For convenience in characterizing the varying depth over the field of MIDIS, we define regions of approximately equal depth:
A (deepest, yellow) defined by the overlapping regions of the MIRI imaging field formed by Obs 1-2, 4 and 5-6, where the exposure time is $37-41.3$h.
B (deep, green) where the exposure time is $\ge 24$h. 
This also includes the Lyot region for Obs 1-2 \& 5-6 and with partial overlap from Obs 4.
The depth in the Lyot region is non-uniform due to the occulting  mask and supporting instrument struts. 
C (NE extension, purple) mainly represents the area of Obs 4 which does not have overlapping observations from Obs 1-2, 5-6, and where the total exposure time amounts to $\ge7$h. In the area outside of these regions (the 'external' region) the depth varies due to partially overlapping dither positions and ranges from $\sim1$ (very small area) to $\sim6$ hours.

The NIRCam and NIRISS parallel observations will be described in forthcoming papers. Some results from the NIRCam parallels have already be published \citep{perez23,caputi24} and more are in preparation.

\begin{table*}
\caption{JWST observations for program 1283}
\label{tab:obs}
\begin{tabular}{c c c c c c c r c l}
\hline\hline
Obs id& PA & dith & offset x,y & nexp & nint & totint & time~~ & date & parallel observation \\
& ($^\circ$) & pos & (arcsec) & & & & (s)~~~& \\
\hline
1 & 27.8 &  1-10 & 0, 0 & 10 & 10 & 100 & 28\,000.2 & Dec 5, 2022 &NIRCam/F115W/F277W \\ \\
2 & 27.8 &  11-20 & 0.0275, 0.03663 & 10 & 10 & 100 & 28\,000.2 & Dec 6, 2022 &  NIRCam/F115W/F277W \\ \\
4 & 36.8 &  31-40 & 10.03663, -32.96337 & 10 & 10 & 100 & 28\,000.2 & Dec 20, 2022 & NIRCam/F150W/F356W \\ \\
5 & 25.8 &  0 & 0, 0  & 1 & 1 & 1 & 277.5 & Dec 2, 2022 & NIRISS/F115W \\
 &  &  41-49 &   & 9 & 4 & 36 & 10\,065.0 & &NIRISS/F115W/GR150R \\
 &  &  0 &   & 1 & 1 & 1 & 277.5 && NIRISS/F115W \\
 &  &  0 &  & 1 & 1 & 1 & 277.5 && NIRISS/F150W \\
 &  &  41-49 &   & 9 & 4 & 36 & 10\,065.0 && NIRISS/F150W/GR150R \\
 &  &  0 &   & 1 & 1 & 1 & 277.5 && NIRISS/F150W \\
 &  &  0 &   & 1 & 2 & 2 & 557.8 && NIRISS/F200W \\
 &  &  41-49 &   & 9 & 4 & 36 & 10\,065.0 && NIRISS/F200W/GR150R \\
 &  &  0 &   & 1 & 2 & 2 & 557.8 && NIRISS/F200W \\ \\
6 & 25.8 &  0 & 0.0328, 0.0328  & 1 & 1 & 1&  277.5 & Dec 3, 2022 &  NIRISS/F115W \\
 &  &  50-58 &   & 9 & 4 & 36 & 10\,065.0 & &  NIRISS/F115W/GR150C \\
 &  &  0 &  & 1 & 1 & 1 & 277.5 & &NIRISS/F115W \\
 &  &  0 &   & 1 & 1 & 1 & 277.5 & & NIRISS/F150W \\
 &  &  50-58 &   & 9 &  4 & 36 & 10\,065.0 & & NIRISS/F150W/GR150C \\
 &  &  0 &   & 1 & 1 & 1 & 277.5 && NIRISS/F150W \\
 &  &  0 &   & 1 & 2 & 2 & 557.8 && NIRISS/F200W \\
 &  &  50-58 &   & 9 & 4 & 36 & 10\,065.0& & NIRISS/F200W/GR150C \\
 &  &  0 &   & 1 & 2 & 2 & 557.8 && NIRISS/F200W \\
 \hline \\
Sum & & & & 96 &  & 532 & 148\,842 \\ 
\\
\hline \\
  \emph{(3} &  &  \emph{21-30} &  & \emph{10} & \emph{10} & \emph{100} & \emph{28\,000.2} & \emph{not executed} & \emph{NIRCam/F150W/F356W}) \\
\hline
\end{tabular}
\tablefoot{Offset with respect to nominal position, applied by SPECIAL REQUIREMENT.
'dith pos' correspond to the points in the MIRI-CYCLING-MEDIUM pattern, and 0 corresponds to the nominal position for no dither. 'nexp' refer to the number of dithered exposures, 'nint' the number of 100 groups integrations obtained at each dither position, and 'totint' the total number of 100 group integrations. Observation 3 could not be executed in 2023 due to repeated safing events of JWST.}
\end{table*}

\begin{figure}
   \centering
  \includegraphics[scale=0.5,trim=0cm 0 0 0]{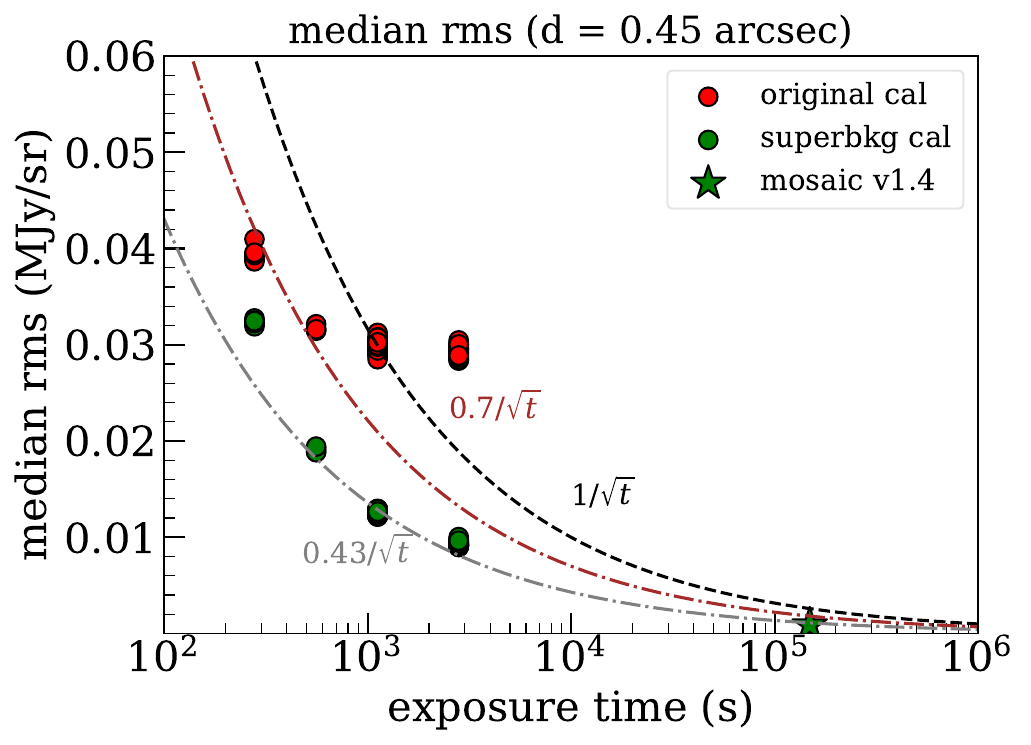}
   \caption{Sky noise vs exposure time for the JWST pipeline output (red) and our customised pipeline including a super-background calibration (green). Note that these have not been corrected for correlated noise induced by the drizzling procedure. The curves show different scalings of the RMS vs exposure time, but are not fits to the data.}
              \label{rms-exptime}%
\end{figure}

\begin{figure*}
   \centering
   \includegraphics[width=0.99\textwidth]{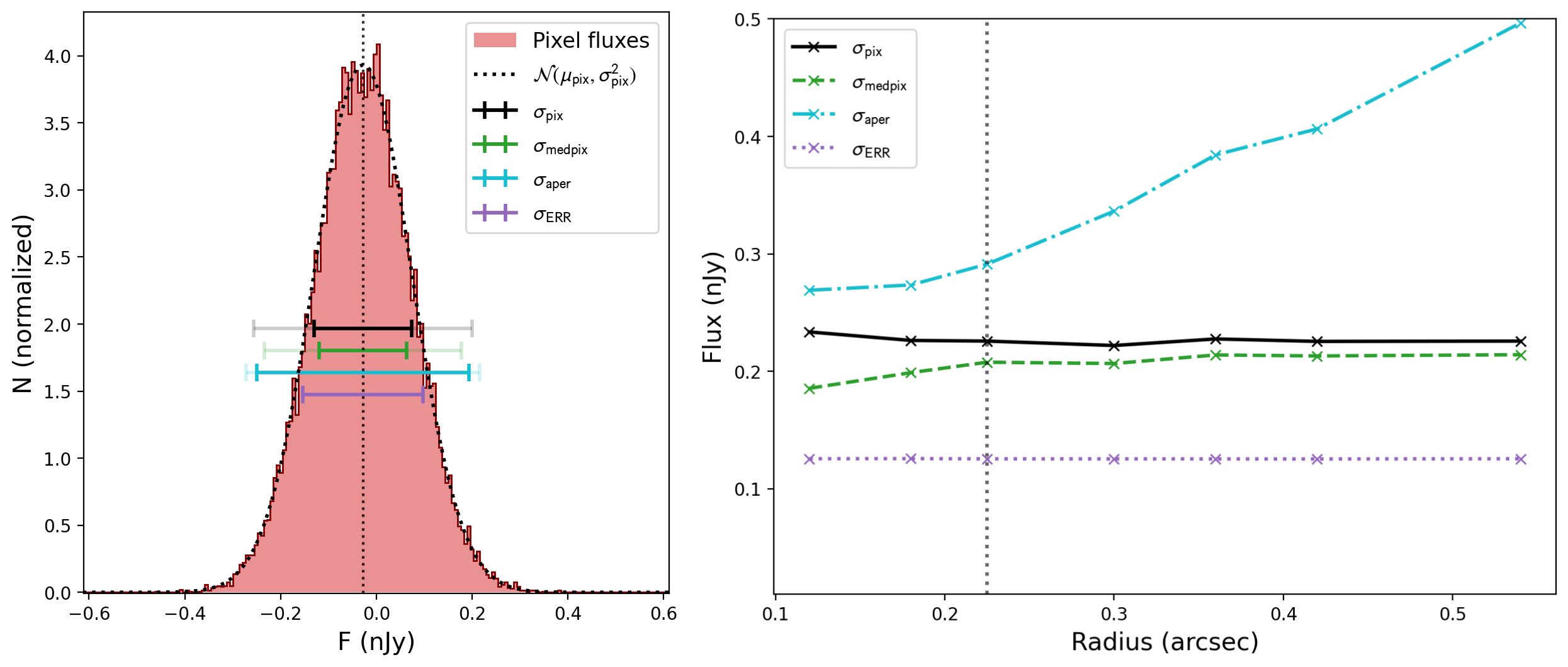}

    \caption{\emph{Left panel:} Flux distribution of all pixels within the sky apertures (see text). The distribution closely follows a normal distribution, showcasing that our method of selecting sky apertures is robust. The black error bar shows the observed standard deviation of the distribution, and the other bars show different noise estimates. The shaded extensions to the bars show the effect of applying the drizzle correlation factor ($R$ and $R_{ap}$, for $\sigma_{medpix}$ and $\sigma_{aper}$, respectively). Comparing the extended green bar to the extended cyan bar then give very similar results. The
black bar, gives the total statistics for all the pixels contained within our apertures. The purple bar shows the noise as inferred from the image error extension ($\sigma_{ERR}$).  \emph{Right panel:} The effect of aperture size on the noise estimates. We use an aperture radius of 0\farcs225 for the final noise estimate which is indicated by the dashed black vertical line. The noise estimates in this panel are all corrected for correlation (except for $\sigma_{ERR}$).}
              \label{noise}
\end{figure*}


\section{Reductions and calibration}

Three different groups within the team made independent reductions to allow for cross-validation of the final images and photometry.
The data were first reduced and calibrated with the JWST pipeline.  This produced images with residual stripes  and residual background \citep{morrison23, dicken24}. We adopted different schemes designed to counter these effects while not distorting the photometric calibration (defining our {\it customised} pipeline, see below).
In addition, we applied an iterative masking of sources for the background estimation: sources appearing after the first run were masked, and then the background was refitted masking new sources that appeared; in total 3 passes were done. 

After careful vetting on the resulting mosaics from the 3 groups, they showed a very good overall photometric agreement, and were found to be largely equivalent. We adopted the version giving the most homogeneous background for our first internal release (v1.4), described below.

The adopted mosaic was reduced with the  {\sc jwst} pipeline version 1.12.3, pmap 1137, with several customised reduction steps especially developed for these data, explained in detail in \citet{perez24a}, and references therein. Briefly, the MIRI F560W data present several artefacts that affect the homogeneity of the background, including a vertical (and a dimmer horizontal) striping and diffuse localized emission linked to cosmic ray showers. To cope with the former, after stage 1 and stage 2 of the JWST pipeline, a master super-background image was constructed for each one of the 96 frames taken by MIDIS, averaging the rest of the images for a given one to be reduced. The background patterns were found to be variable, so we restricted the data to be used to correct a given frame to those taken within a week. For the MIDIS dataset, this translated to completely independent super-background frames for the 2-6 December (Obs 1-2,5-6) and 20 December (Obs 4) datasets. Before the average, sources detected in an initial run of the JWST pipeline were masked, and all frames were homogenized to the same median background of the image being corrected for backgroundinhomogeneity. The method also included a row and column median-filtering and a smooth ($100\times100$ pixel$^2$) background subtraction. These additional processing steps applied to the JWST pipeline defines our {\it customised} pipeline. The effect of cosmic ray showers was partially taken into account by the JWST pipeline, but they are not totally corrected and this is still a significant contributor to the noise in the final mosaic. After the background was homogenized, we calibrated the WCS of each calibrated image using the {\it tweakreg} external routine provided by the CEERS collaboration \citep{bagley23}, using the Hubble Legacy Field catalog \citep{whitaker19} as the reference. Finally, all images were stacked using the JWST pipeline's stage 3. 

The final output image was drizzled to different pixel scales: 
0.11\arcsec\ (nominal MIRI pixel scale), 0.06\arcsec\ for robust MIRI only applications, and also 0.03\arcsec\ and 0.04\arcsec\, for various comparisons with NIRCam data. For the the finer pixel scale mosaics, we note that they may suffer from insufficient sampling in the external region.
In Fig. \ref{image-exposuremap} we show the resulting F560W image and the corresponding exposure map.

\begin{table}
\caption{Summary of noise and depth measurements}
\label{depthtable}
\begin{tabular}{c c c c c c c }
\hline\hline
Area id & Area & Exp time & $\sigma_{\rm medpix}$ & $5\sigma$  limit   \\
& $\Box\arcmin$ & (h) & (nJy/pix) & (AB mag / nJy) \\
\hline
A   & 1.12   & $\gtrsim37$ & 0.20 &  28.65 / 12.6  \\ 
B   & 1.64   & $\gtrsim24$ & 0.24 &  28.48 / 14.7 \\ 
C   & 0.80   & $\gtrsim7$  & 0.50  &  27.68 / 30.8 \\ 
A+B & 2.76   & $\gtrsim24$ & 0.22 &  28.58 / 13.4  \\ 
External & 1.15 & $\sim$1-6 & $\sim0.9$ & $\sim27$ / $\sim50$ \\
 \hline 
\end{tabular}
\tablefoot{Areas are indicated in Figure~\ref{image-exposuremap}. 
$\sigma_{\rm medpix}$ calculated within apertures with diameter $\varnothing=0.45\arcsec$, corrected for the drizzling correlation (a multiplicative factor 2.24, see text for details). The limiting magnitude includes the pixel correlation correction and aperture correction.

}
\end{table}

\begin{figure*}
   \centering
  \includegraphics[width=0.99\textwidth]{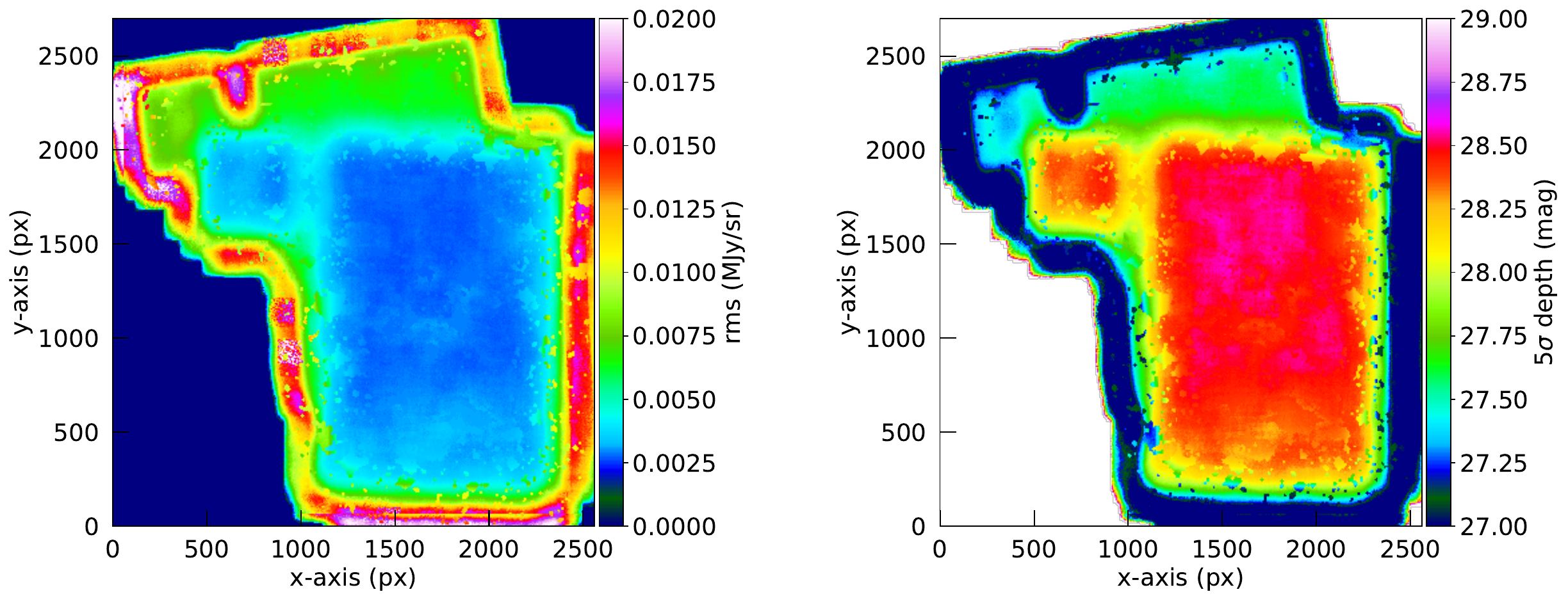}
   \caption{
   RMS (left) and depth (right) map for output scale of 0.06\arcsec/pixel.}
\label{luca-rms+depth}
\end{figure*}

\subsection{Noise measurements and pixel correlation}

In order to determine the depth of the observations, we need to measure the background noise in our images on scales comparable to our detected sources. The noise levels are, in addition to the source size, affected by the photometric zeropoint and an aperture correction which in turn depends on the encircled energy distribution of the PSF. 

In Fig. \ref{rms-exptime} we show how the background sky noise, measured on blank sky in 0.45\arcsec\ diameter apertures,  scales with exposure time for our superbackground calibration (customized pipeline) and the nominal JWST pipeline output, showing the former to scale as $\sim 0.43/\sqrt{t}$, while the latter shows a much more modest decrease of the sky noise with $t$. 
This figure illustrates the improvement in background noise for our data customized pipeline reduction compared to the nominal JWST one.  

The sub-pixel dithering and resampling in the drizzling procedure means that the output pixel fluxes become correlated, which artificially lowers the pixel-to-pixel RMS background noise in the output image (as it would in a box-car smoothing). These effects have been thoroughly discussed in \citet{fruchter02}; and as we have many dither positions uniformly filling the sub-pixel phase space (see Fig. 2), their equations 9 and 10 for calculating the scaling, $R$, with which the RMS should be multiplied, should be directly applicable to our data. 

We produced drizzled image mosaics of different scales ($s=s_{out}/s_{in}$, i.e. output/nominal pixel scale) and {\it pixfrac}  $p\in[1, 0.9, 0.7, 0.5, 0.0]$, enabling us to directly test the effect of pixel correlation on the background noise.  For our nominal image mosaics with pixfrac $p=1$, and scale 0.11\arcsec\ ($s=1$, i.e. native scale), 0.06\arcsec ($s=0.545$), 0.04\arcsec ($s=0.36$), and 0.03\arcsec ($s=0.27$), the noise correction factor that should be applied to the background noise is predicted to be $R=1.5, ~2.24, ~3.13$ and 4.03, respectively. 


There are different ways commonly employed to measure the background/sky noise in reduced images. 
For point sources (or quasi-point sources such as very faint galaxies), it is customary to perform aperture photometry, in an aperture with diameter $d_{ap}$. An aperture placed on blank sky will contain pixel values with a RMS of ${\sigma_{pix}}$. One can then take the average or median of the standard deviations  measured in such apertures as ${\langle\sigma_{pix}\rangle}$;
we here adopt the median, being more robust.

The 1$\sigma$ sky noise, which we denote ${\sigma_{medpix}}$ is then obtained by multiplying 
${\langle\sigma_{pix}\rangle}$ with $\sqrt{N_{pix}}$ (where $N_{pix} = \pi(d_{ap}/2s_{out})^2$  is the number of pixels in the aperture), 
and  the correlation factor $R$  above:

\begin{equation}
\sigma_{medpix} =  {\langle\sigma_{pix}\rangle}\times \sqrt{N_{pix}} \times R 
\end{equation}

For this purpose, we first make a segmentation map, masking out regions with bright emission, and place random apertures in the remaining area, while rejecting those that contain detected faint sources. We do this measurement in 1000 randomly drawn apertures for mosaics generated with different {\it pixfrac} applying the correction ($R$) for correlated noise, and arrive at close to identical estimates for ${\langle\sigma_{pix}\rangle}$. Hence, the formulae of \citet{fruchter02} accurately describes the effect of noise correlation when varying {\it pixfrac}. 

Another commonly used method, that to some extent mitigates the effect of pixel correlation, is to take the standard deviation ({$\mathcal{S}$}) of the total flux $f_{ap}$ in apertures placed on blank sky close to an object: 

\begin{equation}
\sigma_{aper} = \mathcal{S}(f_{ap}) \times R_{ap}
\end{equation}

where ${R_{ap}}$ accounts for the correlation between aperture fluxes \citep[referred to as {\it block sum} in][$R_{ap} \ll R$]{fruchter02}.
For a  flat background of random noise, these two methods are identical. 
However, if the background is not perfectly flat over the image, $\sigma_{aper}$ will be inflated. This  does not affect the $\sigma_{medpix}$ method because that measure uses the median of local noise estimates.

In our final image (v1.4), the low  frequency variations of the background have been removed by the superbackground subtraction, but our tests show that  $\sigma_{aper}$ is lower when measured semi locally (over regions smaller than $10\arcsec\times10$\arcsec) than globally by $\sim 30\%$, while $\sigma_{medpix}$  remains constant, showing that the background subtraction, as expected,  is not perfect and that small/medium spatial scale variations remain. However, when looking at the standard deviation of all pixels within the sky apertures (defined as $\sigma_{pix}$) we find a sky noise which is constant with aperture size and much closer to the $\sigma_{medpix}$ estimate. This reflects that the background variation on pixel to pixel scales is not very strong (but still causing the slightly higher noise level).
The agreement between $\sigma_{pix}$  and $\sigma_{medpix}$ is good for $d_{ap}=0.45\arcsec$, with the former being marginally larger which we attribute to the remaining small scale background variation.  
In Figure \ref{noise} we compare the 
noise measurements described above. The right-hand panel shows that the sky noise
(here per pixel) is roughly constant with aperture size, except for $\sigma_{aper}$.

We finally compared our noise measurements to those provided by the error extension from the pipeline and found that the latter generally underestimates the true background noise\footnote{note that the effects of our custom scripts added to the JWST pipeline have not been fully propagated into the error extension. Since the  JWST pipeline gives larger background noise, the error extension would underestimate the true error even more.}.  The difference is smallest (and negligible) when employing $s=1$ and $p=0$, but raises for smaller pixel scale ($s$) and $p>0$. We estimate the error extension noise ($\sigma_{ERR}$) by measuring the average variance inside the sky apertures.

For the case $s<1, p>0$ we find that the ratio $R_{err} = \sigma_{medpix}/\sigma_{ERR}$ by which the error frame underestimates the noise asymptotically increases with  the aperture size, but is largely constant for $r_{ap}\ge 0.225$ (i.e. $\sim$ two native MIRI pixels) and amounts to $R_{err}=1.72$ for 0.06\arcsec\ ($s=0.545, p=1$) and $R_{err}=2.5$ for 0.04\arcsec\ ($s=0.36, p=1$).  Given that $1 < R_{err} < R$, we conclude that the error frames from the JWST pipeline to some extent but not fully account for the pixel correlation introduced by subpixel dithering and drizzling in MIRI, and should not be taken at face value.

\subsection{Photometric depth}
For the output mosaic with output pixel scale 0.06\arcsec\ the measured $5\sigma$ 
average sky noise for $d_{ap}=0.45\arcsec$ is 29.33 mag ($1\sigma$ 31.08) in the deepest area (A). For a point source in a $d_{ap}=0.45\arcsec$ aperture, the encircled energy fraction is 0.53, and hence our sky noise corresponds to a $5\sigma$ limiting point source magnitude of 28.65. However, given the large variations 
of the exposure time, we provide in Table \ref{depthtable} also the depth for areas B (28.5 mag) 
and C (27.7 mag) and the combined A+B area (28.6 mag). For the {\it external} area, the net exposure times vary more, introducing larger variations in depth, but 27 mag can be taken as a representative value for the $5\sigma$ point source limit (see Table \ref{depthtable}).

In Fig. \ref{luca-rms+depth} we show the variation of RMS sky noise (left) and resulting point source limiting magnitude (right). In making these maps, regions with sources have been interpolated over.

This can be compared to the prediction from the exposure time  calculator which currently gives 28.31 mag for a 40h exposure (relevant for area A) 
in low background conditions (and with background measured in a $r=0.5\arcsec - 1.5\arcsec$ annulus). Hence, we are out-performing the ETC by $\sim 0.35$ mag.
It is likely that an improved handling of the cosmic ray showers could improve the depth of the MIRI Deep Survey even  further. 

Finally, we note that the background noise and point source limiting magnitudes also depend on the pixel scale adopted. The above measurements are for 0.06\arcsec\ and the same values are found for 0.11\arcsec\ (native MIRI pixel scale), while for the 0.04\arcsec\ and 0.03\arcsec\ scales, the images are $\sim0.05$ magnitudes shallower.



\begin{figure}
   \centering
  \includegraphics[scale=0.35]{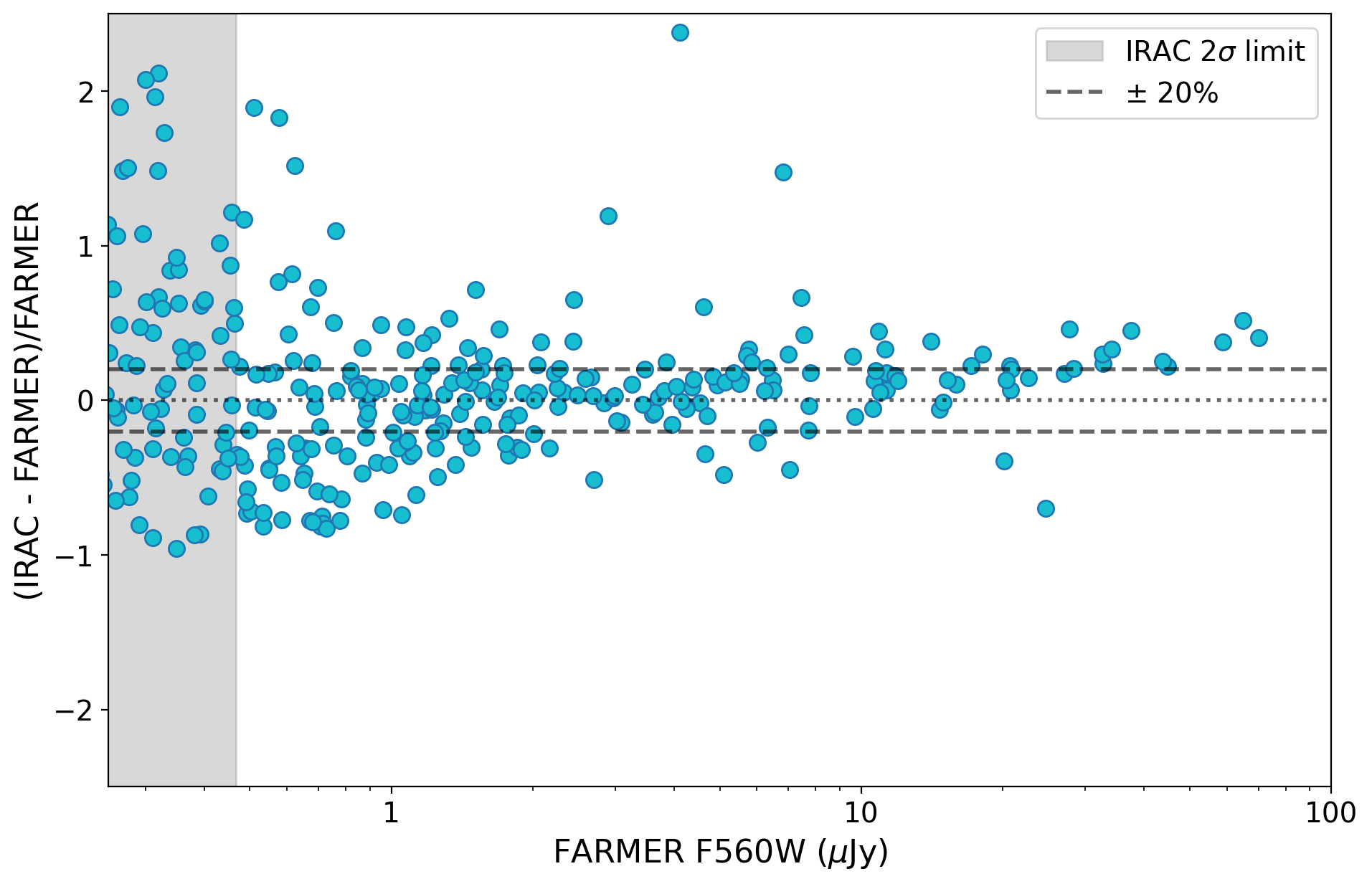}
   \caption{Comparison of MIRI/F560W to IRAC/Ch3 photometry. For $ < 15 \mu$Jy the fluxes agree, while for $>20\mu$Jy the IRAC photometry is systematically brighter. }
              \label{spitzer-vs-miri}%
    \end{figure}

 \begin{figure*}
   \centering
  \includegraphics[scale=0.56, clip=true,trim=2cm 0cm 0 0]{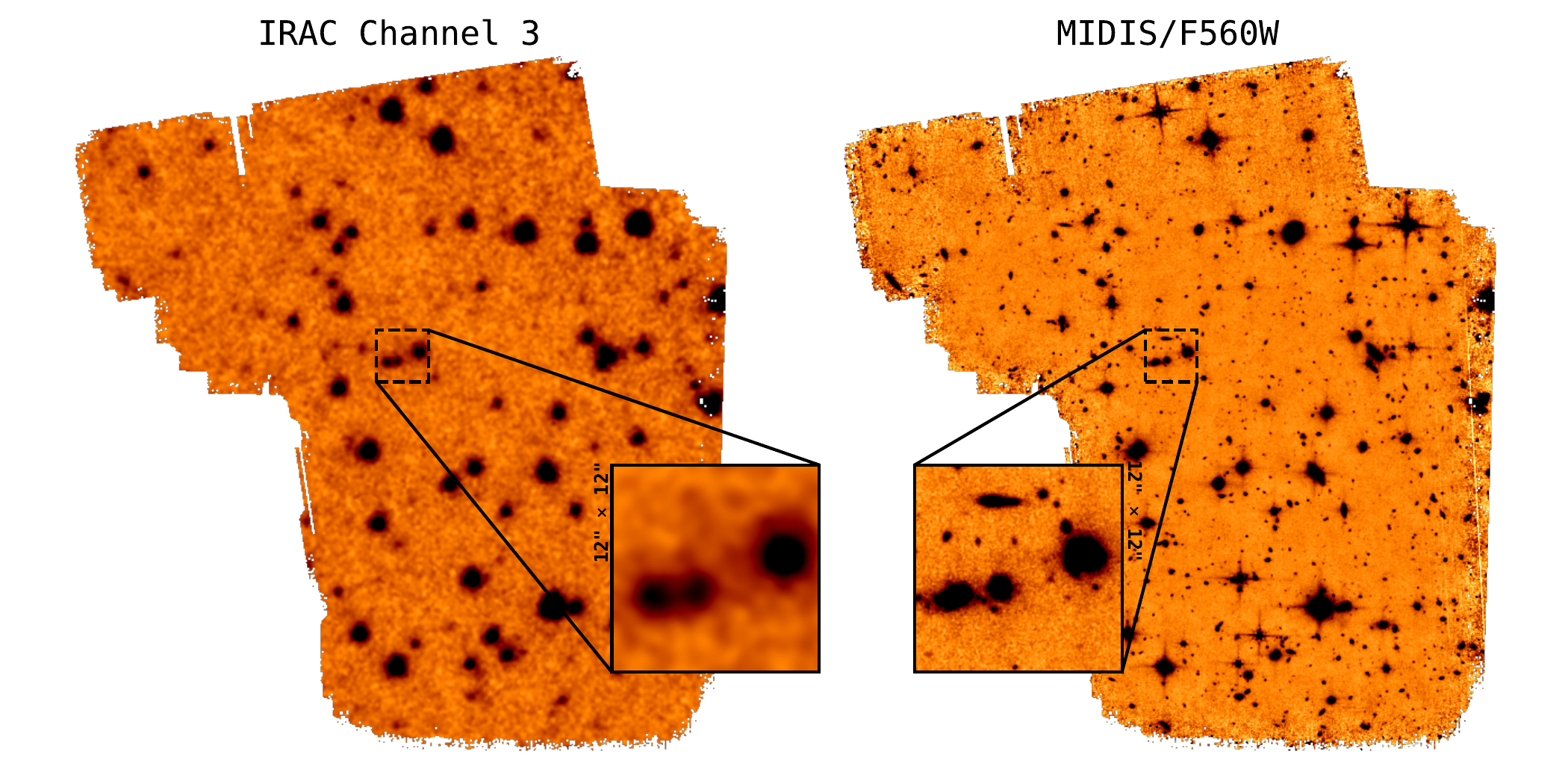}
   \caption{left: Spitzer/IRAC band 3 ($\lambda_{\rm eff}=5.8\mu$m), based on exposure time of $\sim40$h, cropped to the size of the MIDIS field. right: our MIRI/F560W image . 
   The insets show a zoom-in on a $12\arcsec \times 12\arcsec$ region highlighting the improvement in sensitivity and resolution. }
              \label{spitzer}%
    \end{figure*}

\section{Ancillary data}\label{AncData}
For a characterization of the sources we make use of ancillary data of the HUDF/XDF area from HST \citep{illingworth13}, JWST: JEMS \citep{williams23}, FRESCO \citep{oesch23} and JADES \citep{rieke23}. In addition we made use of published Spitzer/IRAC3 photometry \citep{stefanon21}. We also made use of the DAWN JWST Archive (DJA\footnote{https://dawn-cph.github.io/dja/}) that coherently reduces, process and combines all public JWST observations using the {\sc{Grizli}} pipeline \citep{Brammer2024}. To enable multi-wavelength, multi-instrument analysis,  the DJA mosaics are drizzled to a pixel scale of 0.04\arcsec\ and aligned to GAIA DR3 \citep{Valentino2023,Gillman2023,Gillman2024}. In accompanying papers we do, in addition, compare the MIRI results to available data from X-rays (Gillman et al. 2024b, in prep.), ALMA \citep{boogaard24} and MUSE \citep{iani23}.

\section{Photometry and source extraction}
We have performed source detection with \texttt{python/sep} \citep{barbary16} and basic aperture photometry with \texttt{python/photutils} \citep{bradley24}.
In addition we present model based photometry based on {\sc{The Farmer}} \citep{Weaver2023}, which incorporates the effect of a varying PSF FWHM for different filters in HST, NIRCam and MIRI.  

\subsection{Source detection and basic photometry}
\label{detection}
Source detection in the nominal 60 mas F560W image is done using \texttt{sep} \citep{barbary16}, a python wrapper for Source Extractor \citep{bertin96}. In Table~\ref{detparstable} we give the extraction parameters used for the catalogue presented in this paper. We use the error maps from the JWST pipeline (scaled by a factor 1.7 to account for the noise underestimate in these maps, see Section~3.1) multiplied by the \texttt{threshfloat} parameter as the pixel detection threshold. 
As a first control of the detection we apply a a SNR limit of 2 ($2\sigma, ~\Delta$mag<0.5) estimated within the source segment (from the detection segmentation map). With these parameters and SNR rejection we find 2751 sources. Given that this is a single filter detection a large number of the faint sources are expected to be spurious, even when filtering the image before detection. We thus manually inspect the sources, rejecting 170 sources that are obviously spurious (single hot pixels, cosmic ray artefacts, PSF features and some sources near the detector edges) and arrive at a final object list containing 2401 sources. 

\begin{table}
\caption{Source detection parameters }
\label{detparstable}
\begin{tabular}{l l}
\hline\hline
\texttt{sep} Parameter & Value \\
\hline
\texttt{threshfloat} & 0.7 \\
\texttt{minarea} & 9  \\
\texttt{filter\_kernel} & Tophat with radius=0.1\arcsec \\
\texttt{clean\_param}& 1.0 \\
\texttt{deblend\_nthresh} & 128 \\
\texttt{deblend\_cont} & 0.0001 \\
\hline 
\end{tabular}
\end{table}

We estimate the number of spurious sources by running the same detection on an inverted F560W image and applying the same SNR rejection. We also adjust the spurious detection rate for the manual inspection. The rate of spurious detections in our final catalogue is 24\% (of 2571 sources) which is higher than expected given the SNR cut applied.
Hence in addition to the 170 manually rejected sources, there are likely $\sim 400$ spurious sources remaining, many of which will be $\lesssim5\sigma$ sources, and the number of true sources is likely $\approx 2000$.
However, in this first single band catalogue for the deep MIRI data we have elected to be liberal in including sources (both in setting the detection parameters and in the manual inspection). Care should therefore be taken when using this catalogue, and multiple filters (e.g. NIRCAM data or other MIRI filters) should be used for any further characterization of the sources. Note that if we apply a more aggressive SNR cut the spurious rate naturally goes down, at SNR$>$10 the rate is 2\%. Why the spurious detection rate is higher than expected is hard to explain 
but a likely explanation could be image defects that remain in the stacked image due to problems with outlier (in particular cosmic ray) rejection. As the JWST pipeline improves in future releases, it is likely that the number of spurious sources will decrease.  In \citet{cerberus} and Jermann et al. (in prep.) we describe  methods (based on investigating and rejecting individual frames) that can be used to verify that interesting sources are real, even with single filter detection. 
We note that the number of F560W $>2\sigma$ sources with a  counterpart in the JADES catalog is 1879.

Photometry is obtained on the detected sources using \texttt{photutils} and two different apertures: (i) circular apertures with a radius of 0.2\arcsec, and (ii) Kron apertures. While the Kron apertures provides an estimate of the total flux of the sources, the small circular apertures will show smaller statistical errors with the caveat that not all of the flux is included. We aperture correct the circular aperture photometry using our PSF model \citep{boogaard24}. In Fig. \ref{f560err+hist} we display the resulting Kron aperture photometry for MIDIS.

\subsection{Farmer Photometry}

Multi-wavelength (MIRI, NIRCam and HST) photometry is derived for all sources detected in the 0.04\arcsec\ v1.4 F560W image with \textsc{The Farmer} software \citep{Weaver2023}. \textsc{The Farmer} simultaneously models all detected sources using parametric models, allowing  accurate deblending of crowded fields and multi-wavelength datasets with varying resolutions. We opt for the 0.04\arcsec\ F560W image to match with the images available in the DAWN JWST Archive (see Sec \ref{AncData}).

In brief, we first detect and model the galaxies in F560W MIRI band using our F560W PSF model \citep{boogaard24}. We then perform forced photometry in the other multi-wavelength bands, using the \textsc{WebbPSF} models for NIRCam \citep{Perrin2012, Perrin2014} , allowing the flux to vary, whilst keeping the structural parameters fixed (details will be presented in Gillman et al. 2024a, in prep.). 


\subsection{Comparison with Spitzer/IRAC}

We have compared our F560W photometry to that of Spitzer/IRAC channel 3, which has a similar bandpass to F560W, for the HUDF \citep{guo13,stefanon21}. In Figure~\ref{spitzer-vs-miri} we show a comparison of our F560W photometry (measured with FARMER) and IRAC/Ch3. 
For fluxes $F \lesssim 10 \mu$Jy we find very small flux differences 
consistent within the large photometric errors of the Spitzer/IRAC observations.
At bright fluxes $F \gtrsim 20 \mu$Jy, we see a gradual systematic brightening of IRAC vs MIRI, amounting to $\gtrsim 20\%$ at the bright end. Either the Spitzer fluxes are too large, or the MIRI ones too small. The number of sources in the bright end are naturally small, nevertheless we have investigated possible explanations for this offset, which became more apparent for the new MIRI calibration released in September 2023. 

The Spitzer catalog of \citet{guo13} contains matched photometry derived by determining the source position, size, and shape from the HST/F160W filter, matching the source shape to IRAC using the PSF and finding the IRAC flux by fitting the convolved source model to the IRAC data. This is similar to the method used in FARMER, but for our photometry the morphology comes from the F560W image itself.
IRAC/Ch3   samples a factor of 3.5 longer wavelengths than the HST/F160W filter. Most of the bright sources are at low ($z<1$) redshift, meaning that HST/F160W samples $\lambda_{\rm rest} \gtrsim 8000$ \AA\ , while IRAC/Ch3 
samples $\lambda_{\rm rest} \gtrsim 2.9$ \mum . If the size (e.g., the effective radius) or morphology of the source varies with $\lambda$, this may give highly uncertain flux measurements, in particular for sources that are faint in the HST/F160W image, but bright in the mid-IR. In addition, for this redshift range, systematic offsets might be present if galaxies at a given brightness have consistently different morphologies between the rest-frame wavelengths probed by HST/F160W and IRAC/Ch2.
In addition to the FARMER photometry comparison we have also compared the IRAC measurements to photometry from $photutils$ using large Kron apertures which show the same offset in the bright end. A more detailed investigation into this effect is outside the scope of this work, but we note that the MIRI/F560W photometry is obtained directly in high-resolution images and will thus give less model-dependent photometry.

We note that discrepant results for some bright sources were found also for JADES when comparing IRAC/Ch2 to F444W \citep{rieke23}, and in CEERS for MIRI/F560W and F770W \citep{yang23}, and for other MIRI/F560W fields \citep{sajkov24}. At least partly, this can be attributed to source blending in the Spitzer data due to its order of magnitude broader PSF, an effect which is also present in the comparison for our data.

In Figure~\ref{spitzer} we show a comparison of our MIRI/F560W 5.6 \mum\ image to the Spitzer/IRAC/Ch3 5.8$\mu$m image (cropped to the size of the F560W image). The maximum depth of the Spitzer image is $\sim40$ hours reaching 25.6 mag  \citep[at 1$\sigma$, ][hence 5$\sigma$ is $\sim24$ mag]{stefanon21}. The increase in resolution (a factor of $\sim15$) and depth (close to 5 magnitudes) despite the same exposure time is quite apparent.

\begin{figure}
   \centering
  \includegraphics[width=0.95\columnwidth]{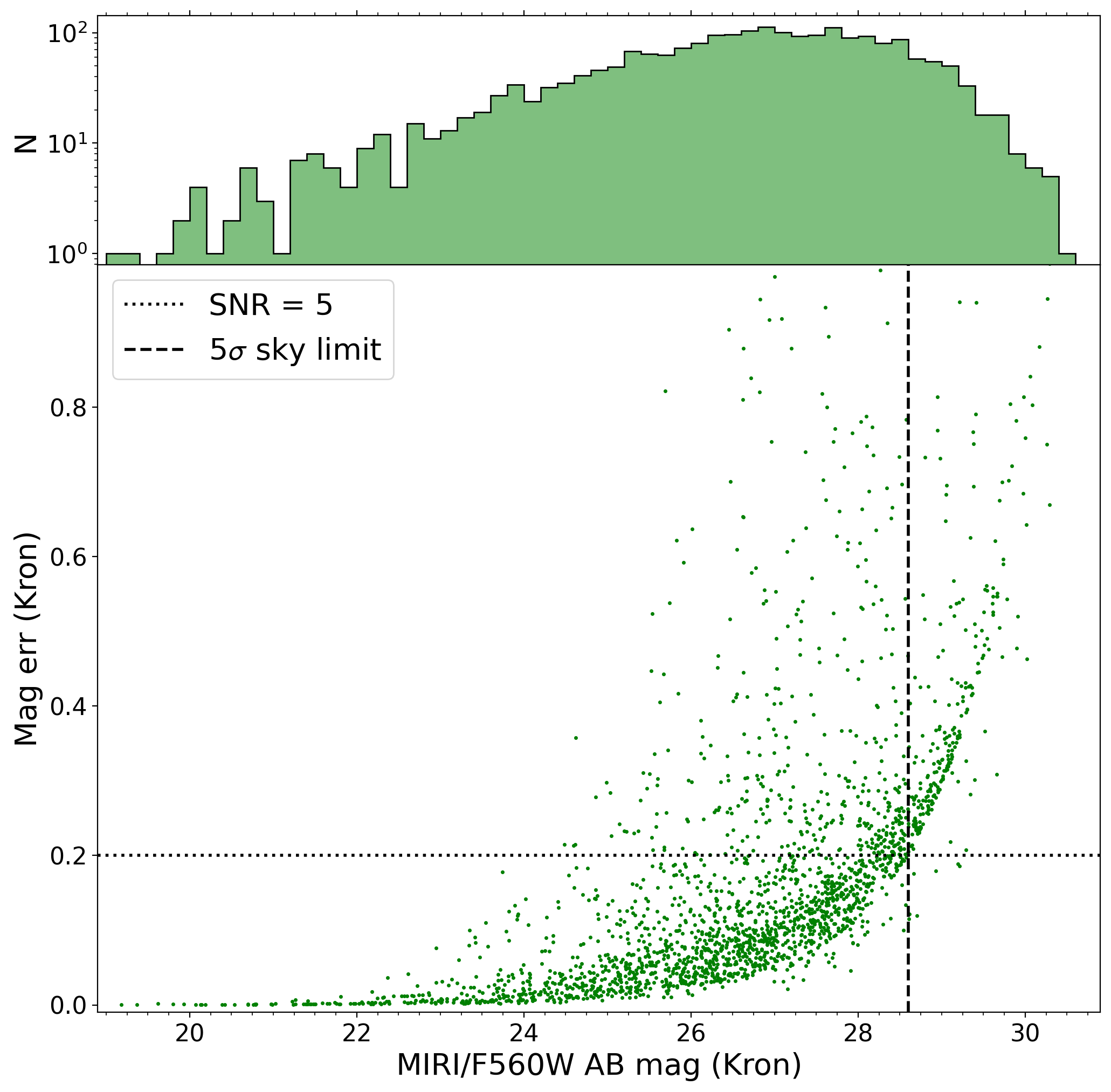}
   \caption{
   Top: F560W number counts for Kron magnitudes.
   Bottom: F560W Kron magnitudes vs photometric uncertainty. The derived $5\sigma$ point source limiting magnitude is indicated by the dashed vertical line. }
\label{f560err+hist}
\end{figure}

\begin{figure}
   \centering
\includegraphics[width=0.95\columnwidth]{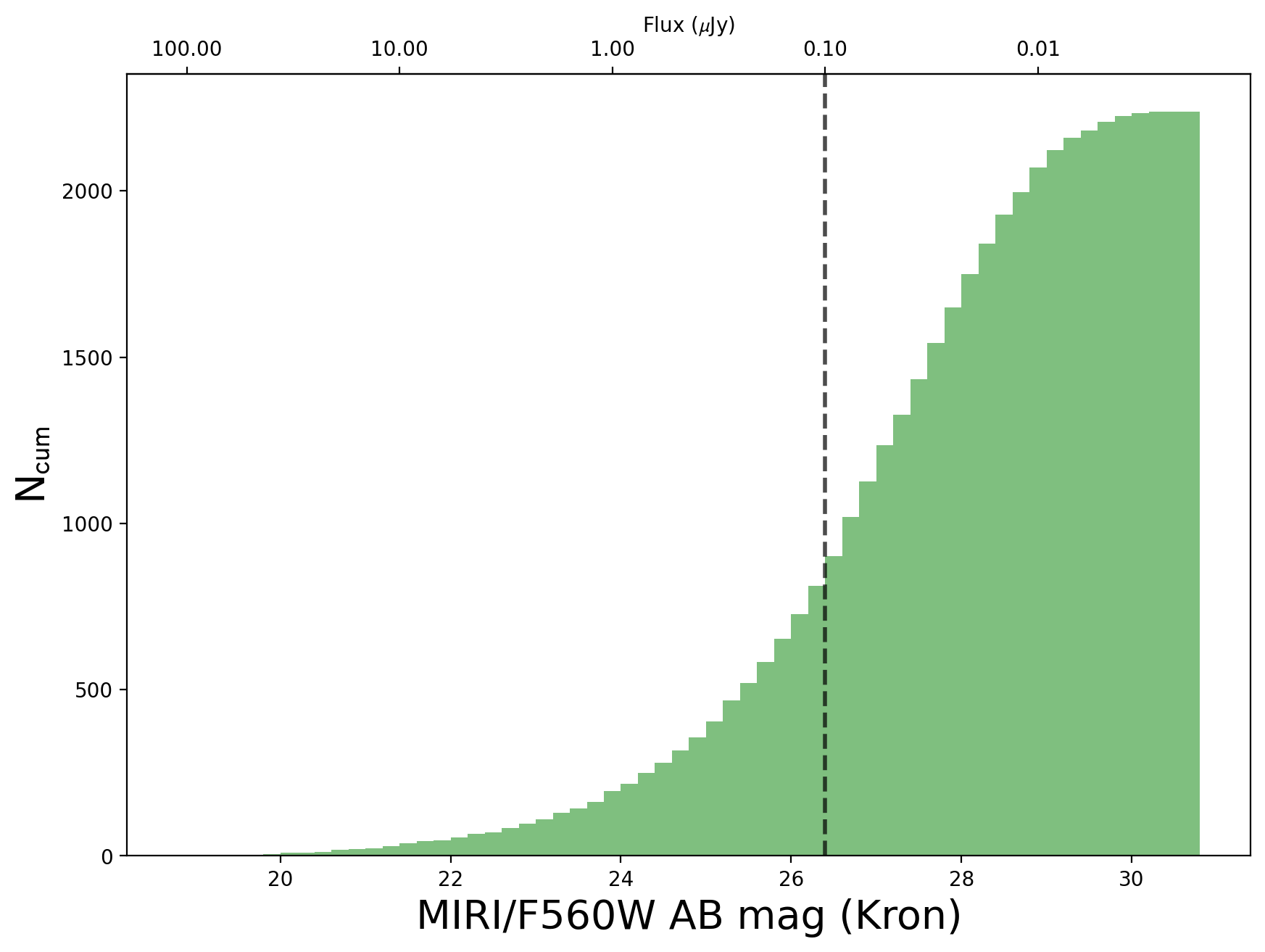}
   \caption{
   F560W cumulative number counts for Kron magnitudes.
    The  dashed vertical line indicates a flux limit of 0.10 \mjy \ as in \citet{sajkov24}. }
\label{f560cumhist}
\end{figure}

\begin{figure*}
   \centering
  \includegraphics[scale=0.8]{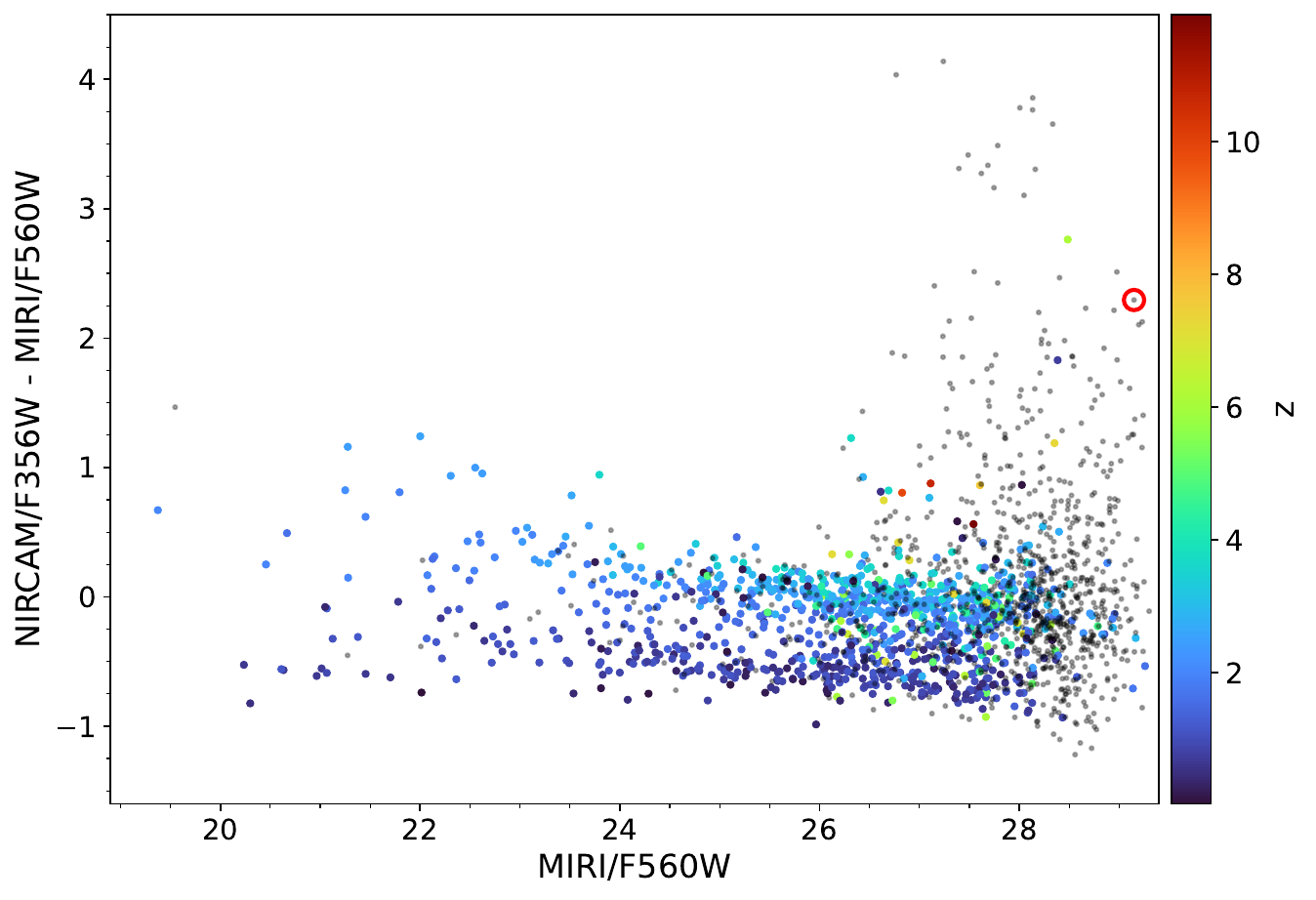}

   \caption{F560W magnitude vs NIRCAM/F356W--F560W color. Only sources with a F560W (F356W) magnitude uncertainty of 0.2 (0.5) or lower are included. The red circle indicates a confirmed MERO currently under investigation (Jermann et al. in prep). The colors indicate photometric redshift of the sources (gray points do not have a valid redshift estimate).}
\label{color-mag}
\end{figure*}

\begin{figure}
   \centering
  \includegraphics[width=0.99\columnwidth]{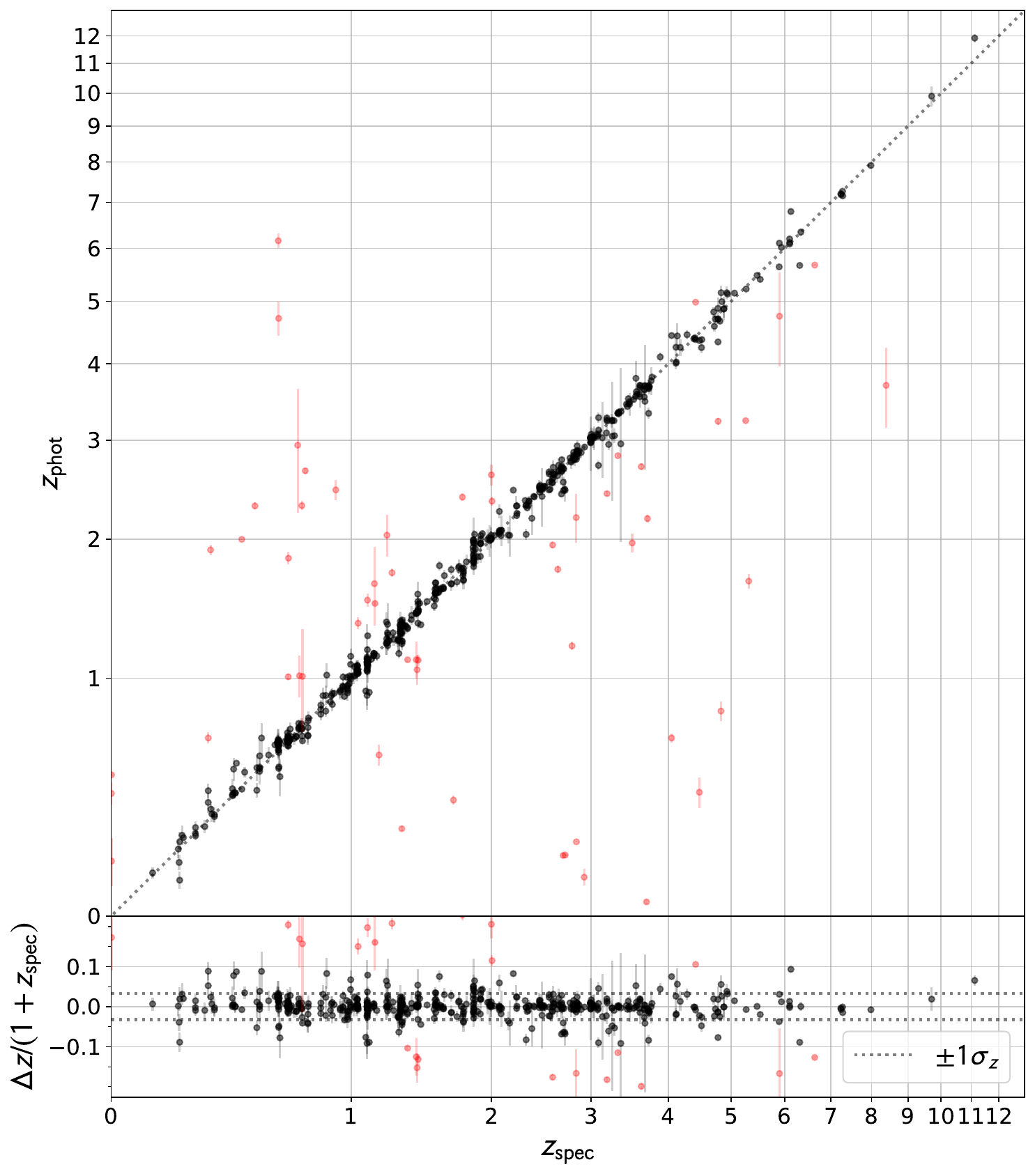}
   \caption{Upper panel: Photometric vs spectroscopic redshifts from the FARMER photometry and the EAZY code. The red points are outliers ($>3\sigma_z$). Lower panel: Normalized difference of the redshifts (with $\Delta z= z_{\mathrm{phot}}-z_{\mathrm{spec}}$) vs spectroscopic redshift. The standard deviation of the normalized difference is $\sigma_z = 0.032$ with the outliers removed.}
   \label{zspec_zphot}
\end{figure}

\begin{figure*}
   \centering
  \includegraphics[scale=0.65
]{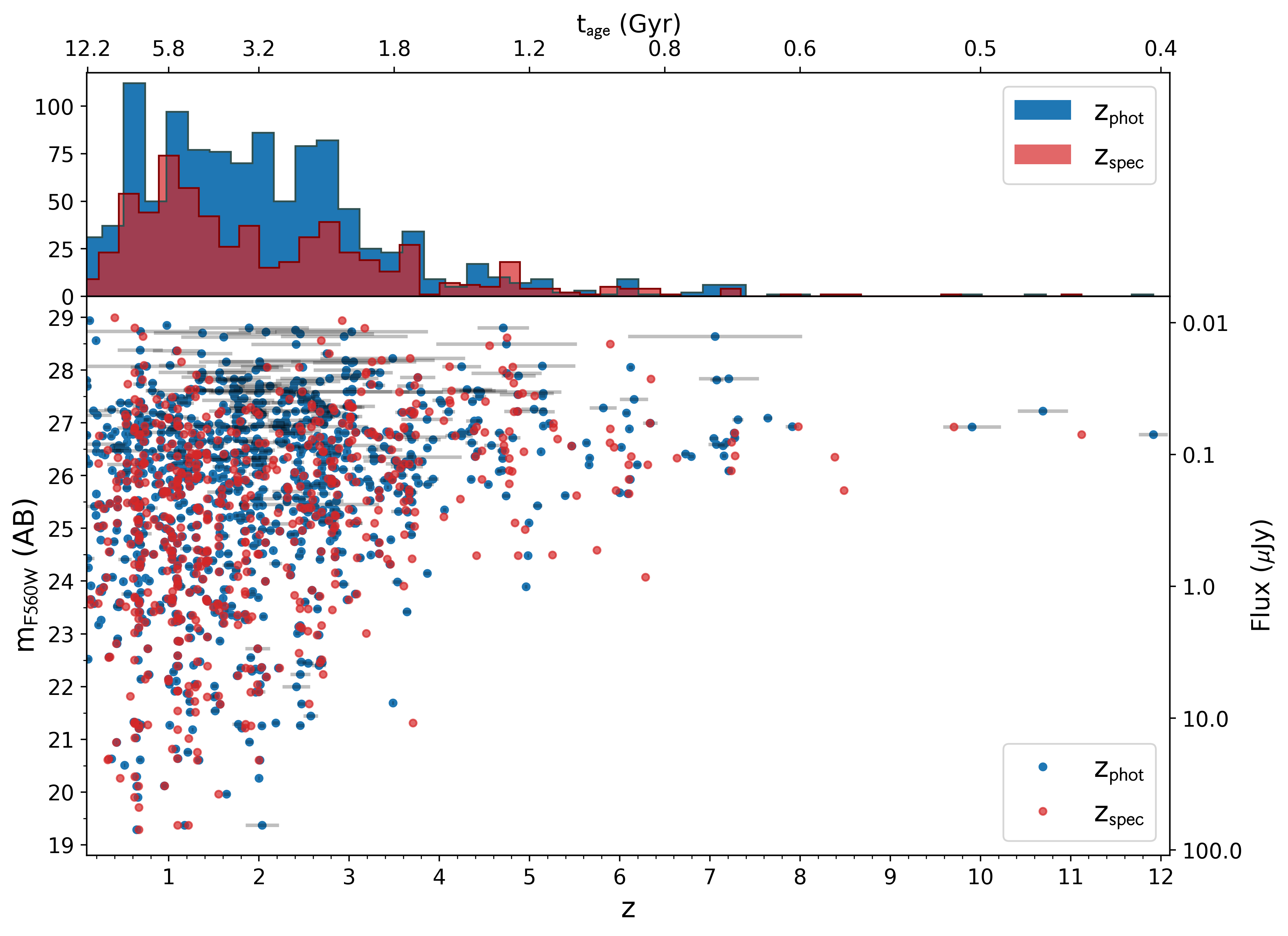}
   \caption{Top: Histogram showing spectroscopic (red) and photometric (blue) redshift distribution. Bottom: Spectroscopic  (red) and photometric (blue) redshift vs F560W magnitude.
   The photometric redshifts are estimated with EAZY. Only sources detected at  $\sigma>2$ in F560W, and with a $1\sigma$ redshift 
   uncertainty $\Delta_z <1$ were included (1069 sources).}
\label{mag-z}
\end{figure*}

\section{Photometric redshifts}

Photometric redshifts were derived from the multi-wavelength {\sc{farmer}} fluxes for each source using the {\sc{eazy}} software \citep{brammer08}. We use thirteen templates from the Flexible Stellar Populations Synthesis code \citep[FSPS;][]{Conroy2010} described in \citet{Kokorev2022} linearly combined to allow for maximum flexibility. We included spectroscopic redshifts from the literature including those from the MUSE Hubble Ultra Deep Survey \citep{Bacon2023}, 3DHST \citep{brammer2012,momcheva2016}, and the JWST Advanced Deep Extragalactic Survey \citep[JADES:][]{Bunker2023,eugenio24}. 

While spectroscopic redshifts have been obtained for many galaxies in the XDF, both before and with JWST, photometric redshifts allow a more complete picture of the z-distribution, and to fainter flux levels. Photometric redshifts based on NIRCam photometry alone are expected to be quite secure, especially for star-forming systems based on the good coverage of the rest frame UV and the Ly$\alpha$ break up to and beyond $z=10$. The power of MIRI is primarily to probe the rest frame optical and near-IR for a more secure determination of the stellar population and identification of intrinsically red, e.g. dusty, sources. Nevertheless it is interesting to investigate how the addition of MIRI affects the phot-z solution.


From the results of {\sc{eazy}} we analysed the systematic offsets between the observed and fitted SED to find a suggested zeropoint (ZP) error of  $\Delta_{ZP} = -0.0388$. This is smaller than the median absolute offset for both NIRCam and HST filters. Hence, the MIRI calibration is from this comparison fully consistent with that of NIRCam, and suggests that the MIRI zero point is accurate. However, when the old (prior to 19 Sept 2023) calibration was used the MIRI offset was instead $\Delta_{ZP} = -0.241$ mag, higher than in any of the NIRCam filters, and suggesting a real ZP offset, similar to the ZP change after 19 Sept 2023. Moreover, for the new calibration, we see no dependence of $\Delta_{ZP}$ for F560W with source brightness, suggesting that the offsets with respect to IRAC/Ch3 likely is due to the IRAC photometry rather than MIRI.

In Fig. \ref{zspec_zphot} we show the comparison of spectroscopic redshifts to photometric redshifts based on {\sc{farmer/eazy}}. In total we find 581 sources with spectroscopic redshifts and F560W photometric uncertainties better than 2$\sigma$ within our field of view. We then define the redshift uncertainty as:
\begin{equation}
\sigma_z = \mathrm{Std}\left(\frac{z_{\mathrm{phot}} - z_{\mathrm{spec}}}{1+z_{\mathrm{spec}}}\right) = \mathrm{Std}\left(\delta z\right), 
\end{equation}
and classify all sources with $\left|\delta z\right| > 3\sigma_z$ as outliers (catastrophic failures). The fraction of catastrophic failures (shown as red circles in the figure) is 10\% and the uncertainty is $\sigma_z = 0.032$. Applying the same method to a comparison of the JADES \citep[data release 2,][]{Eisenstein23} photometric redshifts to the spectroscopic redshifts we find a similar uncertainty and catastrophic failure fraction. As expected the addition of F560W data does not improve the photometric redshift estimates over the full sample (but see Sect. 7.4).

In Fig. \ref{mag-z} we show the redshift distribution vs F560W magnitude. This figure includes, in addition  to the spectro-z's (in red), photometric redshifts (in blue) for all galaxies (1069) for which the photo-z uncertainty $\Delta_z < 1$. Another $\sim$1000 sources have $\Delta_z > 1$, and have $z_{ph}$ distributed from $\sim0$ to $\sim18$, with the majority at $z_{ph}>6$ and with peaks at $z\sim 0.6$, 4.5 and 12. While the individual redshifts of those are not to be regarded as uncertain, a significant fraction is likely to be at 
$z>6$. 
Finally, there are $\sim$500 with no phot-z solution at all from EAZY,
many of which are likely to be spurious sources (see section 5.1).

\section{Results}
The final MIRI image is shown in Fig. \ref{rgb}. As described above, the photometric analysis shows that we reach a 5$\sigma$ point source depth of 28.6 mag (AB), and that this is significantly deeper than the pre-launch and ETC estimates ($\sim 28.3$). This demonstrates the spectacularly successful realisation of JWST \citep{gardner23} and MIRI \citep{wright23}.

We show in Fig.~\ref{f560err+hist}
the resulting F560W magnitude distribution for the 2401 sources (but we note that there are likely spurious sources remaining, especially at $>28$ mag, see sect. 5.1). In Fig.~\ref{color-mag} we show the F560W mag vs F356W--F560W colour for matched sources with $\ge5\sigma$ in F560W and $\ge2\sigma$ in F356W from JADES. For $1.5 < z < 10$, F560W probes the rest frame K-band  (2.2\mum ) to visual (0.5\mum ) while NIRCam/F356W probes the 
rest frame $\sim$H-band (1.6 \mum) to the $\sim$U band (0.32 \mum ). Hence this colour gives an indication of diversity of stellar populations in the sample. Since sources are detected in both filters, and visually inspected, the spurious fraction should be very small, though likely not null.

\subsection{Number counts}
Fig. \ref{f560err+hist} (upper panel)  shows the number counts derived from the total MIRI image (see Fig. 3). The histogram peaks at $\sim27$ 
mag where it is rather flat and then decreases at $>27.5$ mag 
due to increasing incompleteness (varying over areas of different depth, see Fig. 3 and Table 2). Our peak counts at $\sim27$ 
mag are $\sim90$ mag$^{-1}$ arcmin$^{-2}$ 
($3.4\cdot10^5$ per magnitude and $\Box^\circ$), which can be compared to 
Spitzer/Irac/Ch3 finding a factor of $\sim15$ lower surface density at $\sim21$ magnitudes \citep[][converted from Vega to AB magnitudes]{fazio04b}.  Our counts at the same brightness suffer from low numbers given the small area of MIDIS but are consistent with that number.

For comparison, the 
F560W counts in CEERS \citep{yang23} and SMILES \citep{stone24} extend to $\sim26$ and
$\sim 25.6$ magnitudes, respectively.
Deeper F560W counts were presented by \citet{sajkov24}, based on 8 fields with exposure depth of 1.85h per field, reaching $\sim 100\%$ completeness at 0.1 \mjy\ (26.4 mag) and a surface density of (N>0.09~\mjy)$=5.1\pm0.6 \cdot10^{5}$ per square degree. This brightness limit is brighter than what we get in the external area (outside of region C) where the exposure time is between $\sim0.8$ (one dither position, and small fraction of the external area) and 6  hours, and where the $5\sigma$ photometric limit is typically $\gtrsim27$  
or $>0.05$ \mjy\ (see Fig. 7). Hence our total MIDIS area should be representative to 0.1 \mjy\ at $\gtrsim 10 \sigma$ with high completeness and with an insignificant number of spurious sources.   We find (see Fig. \ref{f560cumhist}) 856 sources down to 0.09 \mjy, over a total area of 4.711 square arcmin
or $6.54 \pm 0.22 \cdot10^{5}$ per square degree (where the uncertainty quoted is the 1$\sigma$ Poisson error), hence consistent within 2 $\sigma$
but suggesting that XDF is somewhat overdense at 5.6 \mum. 
The total density of sources with F560W magnitude $\le27$ (0.58 \mjy, where we are reasonably complete over the full area) is $9.6\cdot10^5$ per square degree. 
Pushing the counts to fainter limits ($\lesssim 0.02$\mjy) over areas A+B would require a more careful rejection of spurious sources which can be accomplished by including photometry from JADES and the upcoming deep F770W and F1000W imaging of XDF from JWST GO program 6511. 

\subsection{Redshift distribution}
As mentioned above, and  shown in Fig. \ref{zspec_zphot} for 90\% of the 581 sources with F560W photometry above the 2$\sigma$ level and available spectroscopic redshifts, 
the agreement with the photo-z derived from EAZY is very good. 

In Fig. \ref{mag-z} we show the distribution of spectroscopic and photometric redshifts versus F560W magnitude. For the photo-z's we have only included sources with F560W fluxes $>2\sigma$ and and with a $1\sigma$ redshift uncertainty $\Delta_z <1$, in total 1075 sources. While the selection functions for spectro- and photo-z's are different, their redshift distribution is similar, as shown in the histogram on top.

The bulk of MIRI/F560W sources are at redshifts
$z<4$, but there is still a good number of sources at $4<z<8$. For $z>8$ the distribution thins out, with 3 objects at $z_{phot} \sim 10$ to 12, among them JADES-GS-z11-0 (see Sect. 7.4). These sources all have F560W magnitudes close to 27. There is a large  number of fainter MIRI sources (up to 29th magnitude)  with high redshift ($z_{phot}>6$) solutions but larger uncertainties (and the NIRCam photometric uncertainties are also larger for such, rendering the photo-z 
uncertainty being $\Delta_z >1$), and while uncertain a large fraction of these are expected to be at $z>6$. 

This demonstrates the ability of MIRI to detect galaxies in the EoR, at $z>6$, but requires reaching depths of $>27$ magnitude in F560W. As NIRSPEC campaigns on the HUDF progress, an increased number of MIRI 5.6\mum\ sources with spectroscopic redshifts will likely emerge, and with deeper NIRCam images 
being obtained, also fainter F560W sources will eventually get reliable photo-z's.

\subsection{Morphologies}
\begin{figure}
   \centering
  \includegraphics[width=\columnwidth]{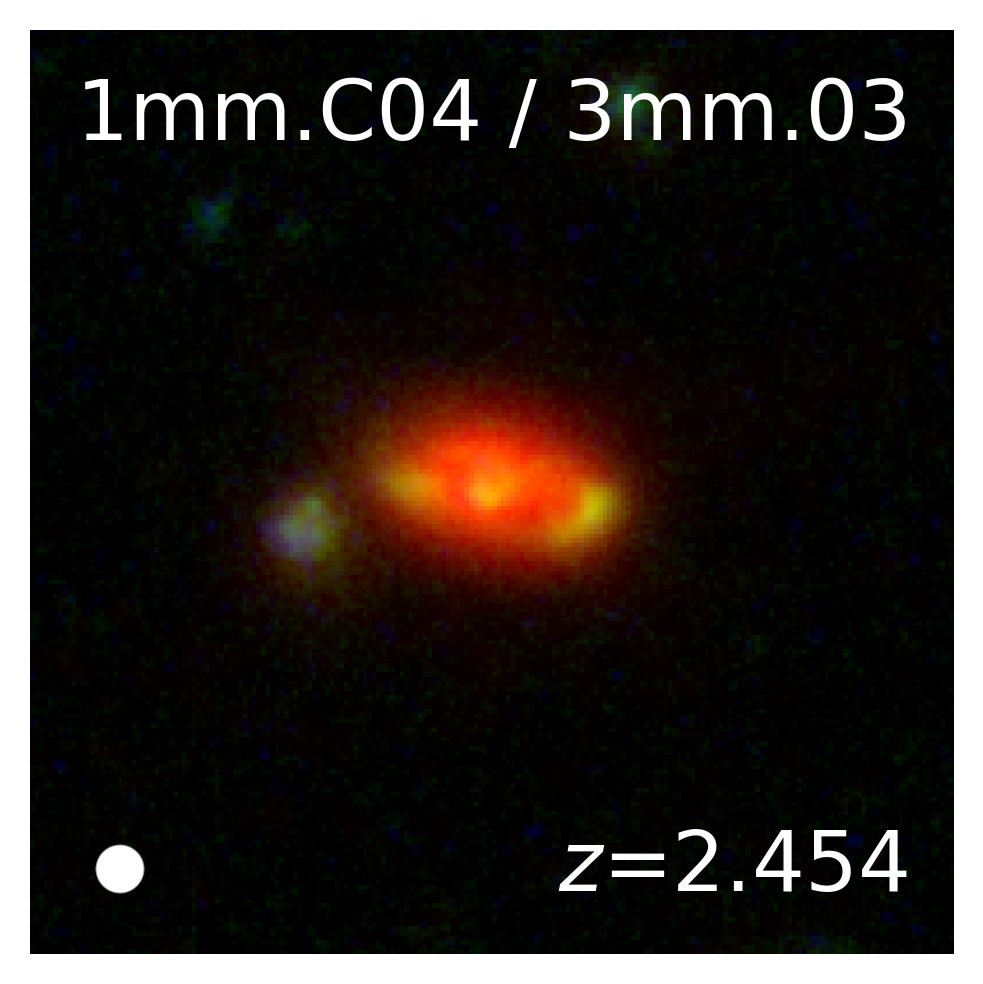}
   \caption{Example of galaxy at cosmic noon ($z\sim2.5$) as seen in MIRI/F560W (red), NIRCam/F182M (green) and HST/F814W (blue) from \cite{boogaard24}. The MIRI emission unveils the presence of a disk-like stellar mass distribution, while the NIRCam  and HST images highlight young star forming regions. The cutout is $4\arcsec\times4\arcsec$ (corresponding to $\sim3\times3$ kpc). The FWHM (0.207\arcsec) of the MIRI/F560W PSF  is shown as a white circle in the lower left. }
\label{morph}
\end{figure}

The FWHM of the MIRI PSF in F560W is $\approx0.2\arcsec$, corresponding to 1.4 kpc at $z=4$. By tracing redder rest frame wavelengths than NIRCam, and with superior resolution and sensitivity compared to Spitzer and WISE, MIRI therefore has a great potential for characterising the stellar mass distribution of high redshift galaxies.
In Fig. \ref{morph} we show an example of a galaxy at z=2.454, also detected in CO and continuum with ALMA \citep{boogaard24} with MIRI in red, 
NIRCam/F182M in green and HST/F814W in blue. 
Here, the MIRI emission (rest frame 1.6 \mum ) is demonstrated to trace the extended stellar light from a more mature stellar population and uniquely shows this to have a disk like structure, while the NIRCam  (restframe visual) and HST (restframe near UV) images highlight young star forming regions \citep{boogaard24}.

Several studies utilising the power of MIDIS to unveil morphology in a more mass (rather than luminosity) weighted manner are on-going (incl. \citealt{costantin24}, Gilman et al. 2024a in prep, Moutard et al. 2024 in prep), as well as a study of X-ray sources and other AGN candidates (Gilman et al. 2024b, in prep).

\subsection{High redshift galaxies detected in MIDIS}
\begin{figure*}
   \centering
   \includegraphics[width=0.99\textwidth]{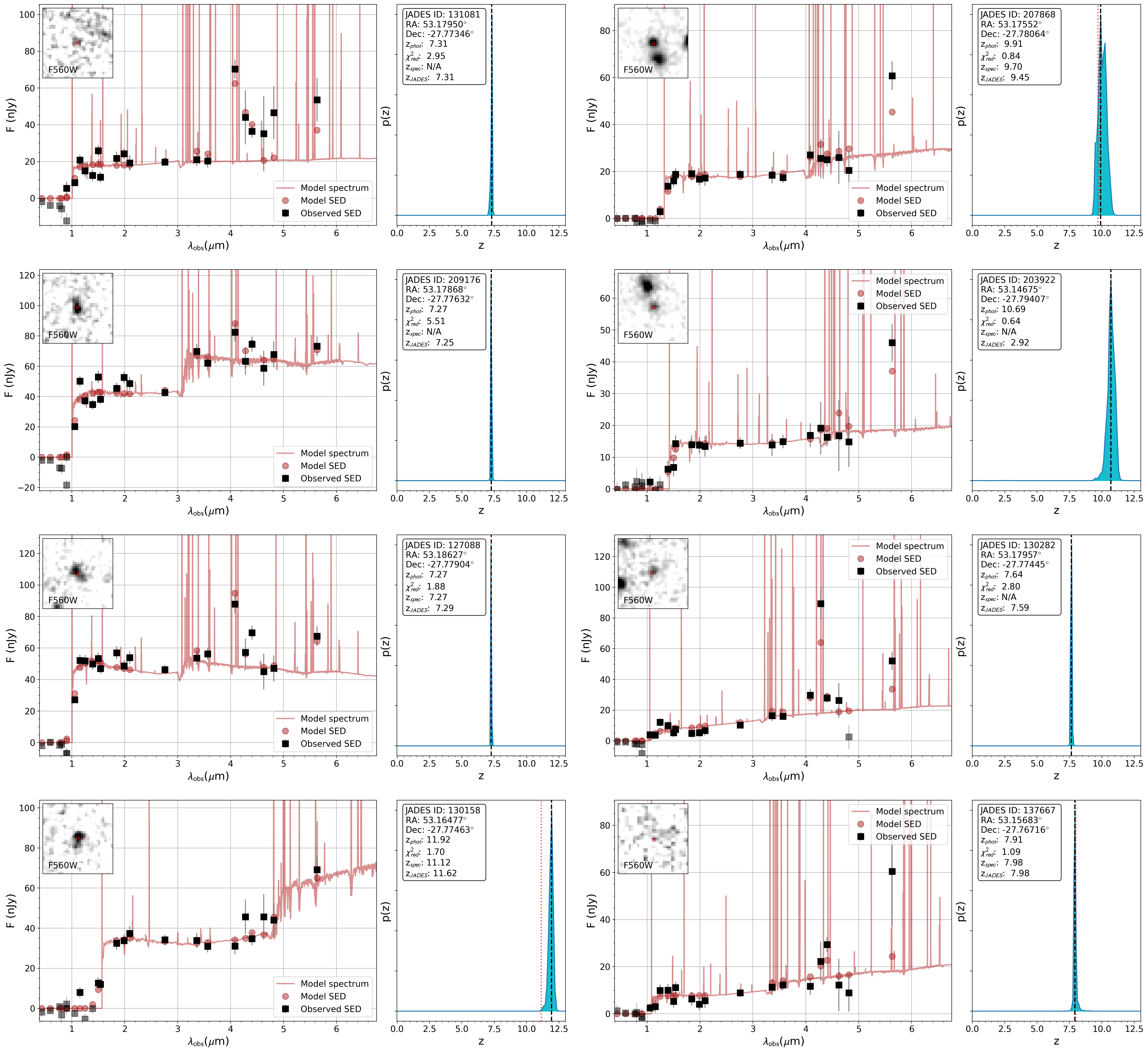}
    \caption{Examples of SED fits made with {\sc eazy}
    for $\gtrsim 8$ galaxies in MIDIS. The insets show F560W $2\arcsec\times2\arcsec$ thumbnails. On the right of each SED we show the redshift probability distribution with insets giving the source identification and redshift estimates. The lower left panel shows JADES-GS-z11-0. The second (from top) panel on the right shows a source for which the addition of MIRI changed the photometric redshift solution from $z=2.9$ to 10.7 due to the bright F560W flux being identified with \hb +\oiii . }
 \label{highz_showcase}
 \end{figure*}

In Figure~\ref{highz_showcase} we show photometric redshift fits of the 8 highest redshift galaxies with reliable photo-z's  in our F560W selected catalogue. The fits are done with EAZY (see Section~6).
The legend contains JADES id number, coordinates, 
JADES photo-z, our photo-z solution and its $\chi^2$, and spectroscopic redshift (available for 4 of the sources). For all but one source (second row, right column) there is a good agreement between the photometric redshifts from JADES and our estimates. The discrepant source has a z=2.92 solution in JADES
(i.e. a the drop in flux shortwards of 1.5\mum\ was identified as a Balmer break), but the bright F560W flux rule out this solution in favor of z=10.7 where
the F560W flux is boosted by \hb +\oiii emission. This illustrates the potential of MIRI for securing and characterizing very high redshift sources, provided faint flux levels can be reached.
In \citet{rinaldi23} we used MIDIS F560W imaging to select 12 strong \ha\ emitters at $z\approx7-8$,
and found them to have high ionising photon production efficiencies \citep[$\xi_{ion}$, see][]{rinaldi24}.

The XDF contains a number of EoR candidates known from past  HST imaging studies \citep[e.g.][including JASDES-GS-z11-0 mentioned below]{oesch13}. The field covered by our F560W image contains two spectroscopically confirmed $z>10$ sources from JADES \citep{robertson23,curtis-lake23}: JADES-GS-z10-0  and JADES-GS-z11-0. While the former is unfortunately contaminated by a F560W PSF spike preventing its secure detection, JADES-GS-z11-0 is well detected in F560W (see Fig. \ref{highz_showcase}, lower left panel, but note that the spectroscopic redshift is somewhat lower than the displayed photo-z fit) for the first time probing the rest frame $>4300$\AA\ emission.  Notably, the galaxy is  extended and presents some structure in the form of a southern extension in F560W, of which there is a hint also in NIRCam images. The extension is treated as a separate source in the JADES catalogue and has a lower (but very uncertain) photometric redshift in that catalogue. This source has recently been proposed to be a companion to JADES-GS-z11-0 \citep{hainline2024}, and our MIRI imaging strengthens this conclusion. The implications of this will be further discussed in a separate letter (Melinder et al., 2024, in prep.).


\subsection{MIRI red sources }
As can be seen from Fig. \ref{color-mag} most sources form a broad sequence of F356W-F560W $\approx-0.15\pm0.6$, but there are also $>100$ sources with much redder colours, $>1$. There is also a similar number of sources with upper ($1\sigma$) limits in F356W, which are in most cases previously unknown (i.e., not catalogued in JADES). Given that the F560W image still likely contains a significant number of spurious sources, sources with upper limits in F356W are not included in Fig. \ref{color-mag} and we do not discuss them further here. Sources with detections in both F560W and F356W are more secure.

We characterise the expected F356W-F560W colours of stellar populations of various age at different redshifts with the Yggdrasil spectral synthesis model\thanks{https://www.astro.uu.se/$\sim$ez/yggdrasil/yggdrasil.html} \citep{yggdrasil} using an instantaneous burst model with metallicity $Z=0.004$ and Kroupa initial mass function.
We find that stellar population ages $<1$ Gyr would for $z\le10$ not be expected to show F560W--F356W colours larger than 1.4,
and for young populations (age $\le 30$Myr) the colour is $<0.1$ for $z<5$ and $<0.5$ for $z\le10$. At z=3 an extinction of $A_V=5$ would add 1.4 magnitude in colour, and at z=8, $A_V=2$ would add 1 magnitude in color \citep[assuming the attenuation law of][]{calzetti00}.  Hence a source with F356W-F560W$>1.5$ could either be a high-z quenched galaxy (with a predominantly old population), or a young (starburst) source with $A_V>5$ at $z\sim3$ or with $A_V>2$
at $z\sim8$. We term such objects 'MIRI extremely red objects' (MEROs).
We note that already F356W-F560W $>1$ would require $A_V>3$ for a $\le30$ Myr stellar population at $z=3$ and $A_V>1$ for $z=8$;  or alternatively a $>300$~Myr old population at $z>8$, and such objects are termed 'MIRI red objects' (MROs). These color criteria do not consider Galactic brown dwarf stars,  which can have very red colors \citep{langeroodi23}, or AGN. Of relevance for the latter class, JWST has discovered a population of red compact sources\citep[.e.g.][]{barro24}, often termed {\em little red dots}, many of which have broad  \ha\ line emission \citep{matthee24}, 
and red NIRCam--MIRI colours \citep{perez24a}, and likely represent a mix of dusty AGN and dusty starbursts.

As evident from Fig. \ref{color-mag}, out of the $\sim60$ MERO candidates identifiable, only 2 sources currently have reliable photometric redshift estimates, which is also true for the similar number of MROs (with $1<$ F356W--F560W$<1.5$).   
However, most of the very red objects are also faint in F560W ($\sim28\pm1.3$ mag). Given the complicated nature of cosmic ray residuals in the MIRI image (see Section~ 5.1) we still suspect that some of these objects could be spurious  (also considering their low S/N in F356W). In order to confirm these objects as real astrophysical sources a more detailed procedure is needed. This investigation is currently underway and will be presented in Jermann at al. (in prep.). One of the confirmed sources is indicated with a red circle in Fig. \ref{color-mag}. Nevertheless, sources with F356W--F560W $>1$ are a potential treasure trove for exploring dusty/old galaxies and AGN at high-z, but requires scrutiny in assessing their reality.

Naturally, the definition of MRO/MEROs according to the above
reasoning  depends on the filter combination used (for instance, for F444W--F560W, the MRO criterion would be $>0.5$), and could be redefined for other color indices, and in selecting sources for further study, all available photometric data should  be used.
Two examples of MROs  identified from other bands (F1000W) in MIDIS are the {\it Cerberus} source \citep{cerberus} detected at 2.8$\sigma$ in F356W, and at 5.9$\sigma$ F1000W (with F356W--F1000W=4.1), but undetected in F560W (upper limit 29.7, implying F356W--F1000W$\lesssim 1.5$); and the AGN candidate {\it Virgil} at z=6.6 \citep{virgil} with F356W--F560W=0.15 (at this redshift the F356W flux is boosted by \hb +\oiii ) but with F444W--F560W=0.85 and F356W--F1000W=1.6. 

The cycle 3 JWST GO project 6511 will image the XDF deeply with MIRI in F770W and F1000W, and will likely produce many more securely identified MIRI red objects.


\section{Data release}
Our F560W image mosaic is made available to the community. Although the data is publicly available in the archive, the image produced by the  {\sc jwst} pipeline version 1.12.3 (pmap 1137) is not yet as optimised as our current reduction (with a customised pipeline). We provide mosaics at pixel scales 0.03, 0.04, 0.06 and 0.11\arcsec . 

We also make a photometric F560W catalog public. The catalog includes both standard photometric properties, and model based photometry from FARMER (when available). 

The images and catalog is available at MAST: {\tt (link will be provided when manuscript is accepted)}. When improvements in the JWST pipeline and calibrations so merits, updated images and catalogs will be uploaded.

\section{Summary and conclusions}

Based on the MIRI European consortium guaranteed time observations, we present MIDIS (the MIRI Deep Imaging Survey) of the Hubble Ultra Deep Field, the deepest 5.6\mum\ image of the universe as yet with a total integration time of 41 hours. We have processed the data with the JWST pipeline  augmented by bespoke adaptations (customised pipeline). 

The deepest parts of the mosaic reach a
point source $5\sigma$ limiting AB magnitude of $28.65$ (12.7 nJy), $\sim 0.35$ magnitudes better than the JWST exposure time calculator predicts.

We detect $\sim$2500 sources down to $2\sigma$. 
Most of these (1879 sources)  have counterparts at shorter wavelengths from JWST/NIRCam and the JADES survey.
Due to remaining issues with the cosmic ray rejection in the JWST pipeline, we expect  $\sim$2000 of the F560W sources to be bona fide distant galaxies.

Comparing the F560W photometry to Spitzer/IRAC Channel 3 observations of the HUDF, we find good agreement but note that for sources brighter than 20.6 mag (20 \mjy ), IRAC fluxes are $\sim25\%$ higher.

There are 581 sources in MIDIS with availble spectroscopic redshifts, and for 1069 sources we could determine accurate photometric redshits (uncertainty in $z<1$). For 90\% of the sources with available spectroscopic redshifts, the agreement with photometric redshifts is excellent ($\sim3\%$).

While the majority of sources with available spectroscopic or accurate photometric redshift are at $z<4$, the redshift distribution extends to $z\sim12$. Many of the sources with less accurate photomtric redshift estimates
are likely to be at $z>6$.

Number counts reveal a total  density of sources $\le 27$ magnitudes ($\sim 0.06$ \mjy) in F560W of 
$\sim10^6$ per square degree. Comparing to
available F560W counts down to 0.09 \mjy\ in
other fields, we find the source density in MIDIS to be $\sim 25\%$ higher.  

MIDIS imaging has been demonstrated to probe the near IR restframe morphology of galaxies out to $z\sim 4$.

More than 100 sources have F356W--F560W colours  $>1$ magnitude, representing MIRI very red sources, suggesting significant dust reddening or old stellar populations
at high redshift.

We illustrate the power of the MIDIS deep 5.6 \mum\ survey by 8 examples of galaxies with $z_{phot}>7$, where in one case the MIRI F560W data point changes the redshift estimate from JADES from $\sim3$ to $>10$.

We make the MIDIS 5.6\mum\ images and a photometry catalog available to the community through MAST.

\begin{acknowledgements}

We dedicate this paper to the memory of our deceased and much valued MIRI-EC team members Hans Ulrik N\o rgaard-Nielsen and Olivier Le F\`evre, both of whom played a central role in defining the MIDIS project.\\

 This work is based on observations made with the NASA/ESA/CSA James Webb Space Telescope.
 The work presented is the effort of the entire MIRI team and the enthusiasm within the MIRI partnership is a significant factor in its success.
 The following National and International Funding Agencies funded and supported the MIRI development: NASA; ESA; Belgian Science Policy Office (BELSPO); Centre Nationale d’Etudes Spatiales (CNES); Danish National Space Centre; Deutsches Zentrum fur Luftund Raumfahrt (DLR); Enterprise Ireland; Ministerio De Economia y Competividad; Netherlands Research School for Astronomy (NOVA); Netherlands Organisation for Scientific Research (NWO); Science and Technology Facilities Council; Swiss Space Office; Swedish National Space Agency (SNSA); and UK Space Agency.
 MIRI drew on the scientific and technical expertise of the following organizations: Ames Research Center, USA; Airbus Defence and Space, UK; CEAIrfu, Saclay, France; Centre Spatial de Li\`ege, Belgium; Consejo Superior de Investigaciones Cientficas, Spain; Carl Zeiss Optronics, Germany; Chalmers University of Technology, Sweden; Danish Space Research Institute, Denmark; Dublin Institute for Advanced Studies, Ireland; European Space Agency, Netherlands; ETCA, Belgium; ETH Zurich, Switzerland; Goddard Space Flight Center, USA; Institute d’Astrophysique Spatiale, France; Instituto Nacional de T\'ecnica Aeroespacial,Spain; Institute for Astronomy, Edinburgh, UK; Jet Propulsion Laboratory, USA; Laboratoire d’Astrophysique de Marseille (LAM), France; Leiden University, Netherlands; Lockheed Advanced Technology Center (USA); NOVA Opt-IR group at Dwingeloo, Netherlands; Northrop Grumman, USA; Max Planck Institut f \"ur Astronomie (MPIA), Heidelberg, Germany; Laboratoire d’Etudes Spatiales et d’Instrumentation en Astrophysique (LESIA), France; Paul Scherrer Institut, Switzerland; Raytheon Vision Systems, USA; RUAG Aerospace, Switzerland; Rutherford Appleton Laboratory (RAL Space), UK; Space Telescope Science Institute, USA; Stockholm University, Sweden; Toegepast- Natuurwetenschappelijk Onderzoek (TNOTPD), Netherlands; UK Astronomy Technology Centre, UK; University College London, UK; University of Amsterdam, Netherlands; University of Arizona, USA; University of Cardiff , UK; University of Cologne, Germany; University of Ghent; University of Groningen, Netherlands; University of Leicester, UK; University of Leuven, Belgium;  Utah State University, USA. \\
 
 Additional acknowledgements related to specific grants: G\"O, JM \& AB acknowledges funding from the Swedish National Space Administration (SNSA).  P.G.P.-G. acknowledges support from grant PID2022-139567NB-I00 funded by Spanish Ministerio de Ciencia e Innovaci\'on MCIN/AEI/10.13039/501100011033,
FEDER {\it Una manera de hacer
Europa}. This work was supported by research grants (VIL16599,VIL54489) from VILLUM FONDEN.
L.C. acknowledges support by grant PIB2021-127718NB-100 from the 
Spanish Ministry of Science and Innovation/State Agency of Research 
MCIN/AEI/10.13039/501100011033 and by “ERDF A way of making Europe”.
JPP and TVT acknowledge financial support from the UK Science and Technology Facilities Council, and the UK Space Agency.
AAH acknowledges financial support from grant PID2021-124665NB-I00 
funded by MCIN/AEI/10.13039/501100011033 and by ERDF A way of making 
Europe.
EI and KC acknowledge funding from the Netherlands
Research School for Astronomy (NOVA). KIC acknowledges funding from the Dutch Research Council
(NWO) through the award of the Vici Grant VI.C.212.036.
RAM acknowledges support from the Swiss National Science Foundation (SNSF) through project grant 200020\_207349. \\
 The paper uses JWST data from   program \#1283,  obtained from the Barbara Mikulski Archive for Space Telescopes at the Space Telescope Science Institute (STScI). 

 \end{acknowledgements}

%
%

\end{document}